\documentclass{article}
\usepackage{arxiv}
\usepackage[utf8]{inputenc} % allow utf-8 input
\usepackage[T1]{fontenc}    % use 8-bit T1 fonts
\usepackage{hyperref}       % hyperlinks
\usepackage{url}            % simple URL typesetting
\usepackage{booktabs}       % professional-quality tables
\usepackage{amsfonts}       % blackboard math symbols
\usepackage{amssymb} 
\usepackage{nicefrac}       % compact symbols for 1/2, etc.
\usepackage{microtype}      % microtypography
\usepackage{lipsum}
\usepackage{graphicx}
 \usepackage{amsmath}
\graphicspath{ {./images/} }

\title{Membrane Heterogeneity Driven Dynamics of Multicomponent Vesicles in Shear Flow}

\author{
 Shuqi Tang \\
 Mechanics Division\\
 Beijing Computational Science Research Center\\
 Beijing, 100193, China\\
   \And
 Steven M. Wise \\
 Department of Mathematics\\
 The University of Tennessee\\
 Knoxville, TN 37996, USA\\
  \And
 John Lowengrub \\
 Department of Mathematics\\
 University of California at Irvine\\
 Irvine, California 92697, USA\\
  \And
 Zhenlin Guo \thanks{Corresponding author. Email: \texttt{zguo@csrc.ac.cn}} \\
 Mechanics Division\\
 Beijing Computational Science Research Center\\
 Beijing, 100193, China\\
}

\begin{document}
\maketitle

\begin{abstract}
Despite their significance in biology and materials science, the dynamics of multicomponent vesicles under shear flow remain poorly understood because of their nonlinear and strongly coupled nature, especially regarding the role of membrane heterogeneity in driving nonequilibrium behavior. Here we present a thermodynamically consistent phase-field model, which is validated against experiments, for the quantitative investigation of these dynamics. While prior research has primarily focused on viscosity or bending rigidity contrasts, we demonstrate that surface tension heterogeneity can also trigger swinging and tumbling in vesicles under shear. Additionally, our systematic phase diagram reveals three previously unreported dynamical regimes arising from the interplay between bending rigidity heterogeneity and shear flow. Overall, our model provides a robust framework for understanding multicomponent vesicle dynamics, with findings offering new physical insights and design principles for tunable vesicle-based carriers.
\end{abstract}

%{\bf MSC Codes }  {\it(Optional)} Please enter your MSC Codes here

\section{Introduction} 
\label{sec:introduction}

Multicomponent lipid vesicles-composed of saturated lipids, unsaturated lipids and cholesterol-serve as minimal models for biological membranes. 
Their lateral phase separation into liquid-ordered ($l_o$) and liquid-disordered ($l_d$) domains \cite{baumgart_imaging_2003, veatch_separation_2003, veatch_miscibility_2005, lingwood_lipid_2010, wang2022lipid} leads to spatial heterogeneity in membrane properties such as bending rigidity and surface tension \cite{baumgart_membrane_2005, yanagisawa_shape_2008}. These heterogeneities underlie lipid rafts, which play critical roles in signal transduction \cite{simons_functional_1997, simons2000lipid, sackmann_physics_2014}, membrane trafficking \cite{simons_model_2004}, and molecular sorting \cite{mukherjee_membrane_2004}. Understanding how such heterogeneity modulates membrane deformation under flow is crucial for the design of biomimetic materials \cite{elani_vesicle-based_2014} and drug delivery in cell biology \cite{allen_drug_2004, herrmann_extracellular_2021, karaz_liposomes_2023, kooijmans_exploring_2021}.

The dynamics of single-component vesicles in shear flow have been extensively investigated through experimental, theoretical, and numerical methods, revealing a rich variety of behaviors such as tank treading, tumbling, and swinging, primarily driven by viscosity contrast \cite{keller_motion_1982, misbah_vacillating_2006, vlahovska_dynamics_2007, noguchi_swinging_2007, kantsler_transition_2006, abkarian_dynamics_2005, deschamps_phase_2009,danker_dynamics_2007,lebedev_nearly_2008}. In contrast, the dynamics of multicomponent vesicles under shear flow remain poorly understood. Two-dimensional (2D) simulations have successfully reproduced phase treading and tumbling driven by bending heterogeneity and asymmetric phase distributions \cite{gera_swinging_2022, lowengrub_surface_2007, sohn_dynamics_2010, liu_dynamics_2017}. However, these models are limited to planar dynamics and cannot account for three-dimensional (3D) phenomena, such as vertical ring banding \cite{gera_three-dimensional_2018, gera_modeling_2018}, nor can they incorporate line tension, which experimental studies have shown can induce tumbling \cite{tusch_when_2018}. Recent 3D simulations provide insights into the interplay of bending heterogeneity and flow \cite{venkatesh_shape_2024}, but are limited by assumptions of small deformations and neglect surface tension heterogeneity. Likewise, the two-phase fluid deformable surface sharp-interface model proposed in \cite{sischka_two-phase_2025} incorporates bending heterogeneity but fails to capture topological transitions and to account for surface tension heterogeneity.

To overcome these limitations, we developed a thermodynamically consistent phase-field model for multicomponent vesicles under shear flow. Our model effectively captures the coupling between background fluid and membrane dynamics, together with lateral phase separation, while incorporating line tension and component-dependent physical properties. Our 3D simulations demonstrate excellent qualitative and quantitative agreement with experimental observations, accurately capturing phase separation and shear-induced dynamics. We further explore how shear-induced dynamics are governed by bending rigidity heterogeneity (\(\kappa_B=\kappa_{B_{l_d}}/\kappa_{B_{l_o}}\)), surface tension heterogeneity (\(\sigma_S=\sigma_{S_{l_d}}/\sigma_{S_{l_o}}\)), and  bending capillary number $\text{Ca}=\mu_{{out}}UL^2/\kappa_{B_{l_o}}$. Our results uncover a range of complex dynamic regimes, including three previously unreported ones. Notably, we identify a surface-tension-driven swinging and tumbling mechanism that arises without viscosity contrast or asymmetric phase distributions. These findings reveal previously unexplored physical pathways for regulating vesicle dynamics and offer guidance for biomimetic vesicle design in flow environments.

The outline of the paper is as follows. Section~\ref{model} presents a thermodynamically consistent phase-field model for multicomponent vesicles, and introduce key dimensionless parameters. Section~\ref{sec.3 validation} validates the model against experimental observations \cite{wang2022lipid,tusch_when_2018}. Section~\ref{sec.4 simulations} reports numerical results, elucidating the interplay between membrane heterogeneity and flow-induced forces. The Appendices present additional methodological and numerical details.

% The appendices provide supplementary details: the coupling with the diffuse domain method (Appendix\ref{appendix-ddm}), derivations of the governing equations (Appendix~\ref{appendix-modle}), and the numerical methods employed in the simulations (Appendix~\ref{appendix-numerical method}).

 \section{Model} \label{model}\
 
\begin{figure}[h]
   \centering
   \includegraphics[width=0.8\linewidth]{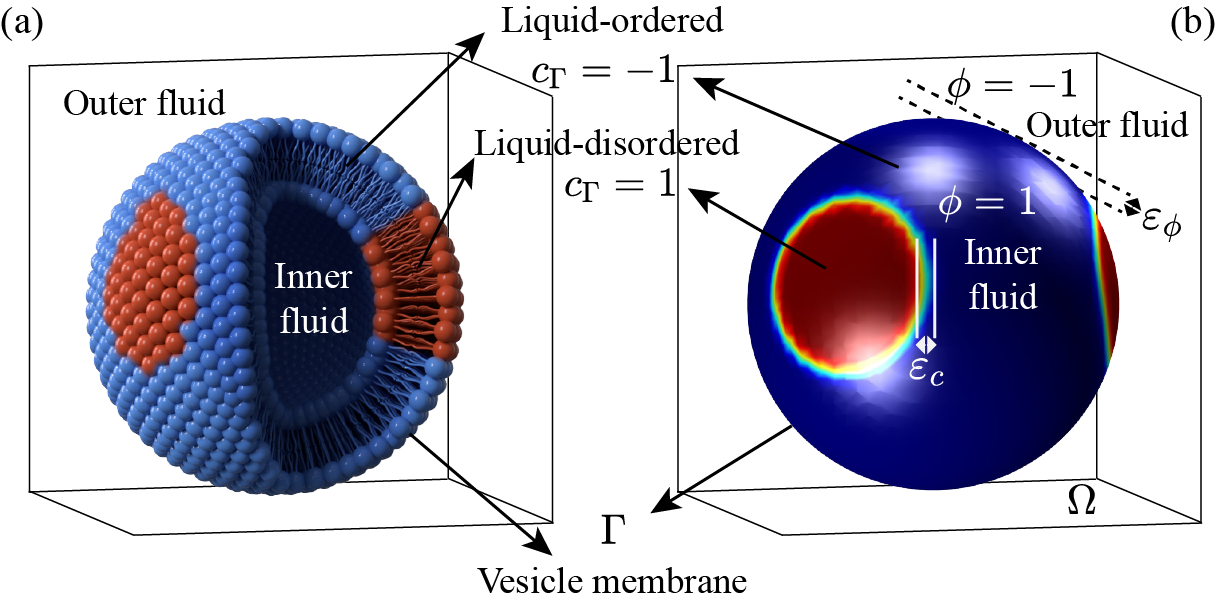}
   \caption{(a) A multicomponent vesicle with lipid components ($l_d$ (blue) and $l_o$ (red)) exhibiting distinct properties, immersed in a background fluid, illustrating the lipid bilayer structure. For simplicity, the two components are distinguished only by color. (b) Diffuse-interface representation of panel (a), showing the corresponding phase-field distribution.}
   \label{fig:schematic}
\end{figure}

In this section, we develop a thermodynamically consistent phase-field model derived from the total energy of the system to investigate the dynamics of multicomponent vesicles with membrane heterogeneity under shear flow within a rectangular domain $\Omega$ (see figure~\ref{fig:schematic}). 

\subsection{Governing equations for vesicle membrane interacting with background fluid}\

We consider a vesicle immersed in a background fluid within a rectangular domain $\Omega$, whose membrane is represented by a closed surface $\Gamma$. The system is modeled as a two-phase flow, with $\Gamma$ serving as the interface between the inner and outer fluids. To capture complex shape deformations, we adopt a phase-field formulation in which the sharp interface $\Gamma$ is replaced by a diffuse interface characterized by a scalar field $\phi(\boldsymbol{x}, t)$. The phase-field variable $\phi$ follows a hyperbolic tangent profile across the membrane interface, ensuring a smooth transition from the interior ($\phi = 1$) to the exterior ($\phi = -1$) of the vesicle. Specifically, $\phi$ can be defined as
\begin{align} \label{eq:phi initial}
\phi(\boldsymbol{x}, t) = \tanh\left(\frac{ r(\boldsymbol{x}, t)}{2\sqrt{2}\varepsilon_\phi}\right), \quad \boldsymbol{x} \in \bar{\Omega},
\end{align}
where $r(\boldsymbol{x}, t)$ denotes the signed distance from $\boldsymbol{x}$ to the nearest point on the membrane interface $\Gamma$, taken positive inside and negative outside, and $\varepsilon_\phi$ controls the thickness of the diffuse interface. 

Both the inner and outer fluids are assumed to be incompressible, with equal density ($\rho_{{in}} = \rho_{{out}}$) but distinct viscosities, denoted by $\mu_{{in}}$ and $\mu_{{out}}$. The governing equations for the coupled two-phase flow and membrane evolution are formulated as follows:
\begin{align}
        \label{ns_un}
	(\rho \boldsymbol{u})_t +  \boldsymbol{\nabla} \cdot  (\rho\boldsymbol{u} \otimes \boldsymbol{u}) &= \boldsymbol{\nabla} \cdot (\mu \boldsymbol{D}(\boldsymbol{u}))  - \boldsymbol{\nabla}  p + \boldsymbol{F} +\boldsymbol{\nabla} \cdot (\zeta(\phi) \boldsymbol{P} \lambda_{{local}}), \\ 
	\label{incompress_un}
	\boldsymbol{\nabla} \cdot \boldsymbol{u}&=0,\\
	\label{ch_un}
	\phi_t+\boldsymbol{\nabla} \cdot(\phi \boldsymbol{u} )&= \boldsymbol{\nabla} \cdot \boldsymbol{q_{\phi}}, \\
     \label{inextend_un}
\xi \varepsilon_\phi^2 \boldsymbol{\nabla} \cdot (\phi^2 \boldsymbol{\nabla} \lambda_{{local}}) + \zeta(\phi) \boldsymbol{P} : \boldsymbol{\nabla} \boldsymbol{u} &= 0,
 \end{align}
where \eqref{ns_un} and \eqref{incompress_un} represent the conservation of momentum and fluid incompressibility via the Navier-Stokes (NS) equations, and \eqref{ch_un} describes the evolution of the vesicle membrane using a phase-field Cahn-Hilliard (CH) formulation. The membrane inextensibility condition is enforced through \eqref{inextend_un}. Here, $\boldsymbol{u}$ and $p$ denote the bulk velocity and pressure, respectively. The rate of strain tensor is given by $\boldsymbol{D}(\boldsymbol{u}) = (\boldsymbol{\nabla} \boldsymbol{u} + \boldsymbol{\nabla} \boldsymbol{u}^{\mathrm{T}})/2$. The spatially varying viscosity is interpolated as $\mu(\phi) = {(1-{\phi})\mu_{in}}/{2} + {(1+{\phi})\mu_{out}}/{2}$. The vesicle-fluid interaction force $\boldsymbol{F}$ and the interfacial flux $\boldsymbol{q}_\phi$ will be defined later.

In addition, membrane inextensibility is modeled using the diffuse domain approach of \cite{aland_diffuse_2014}, which extends the local surface constraint $\boldsymbol{\nabla}_{\Gamma} \cdot \boldsymbol{u} = \boldsymbol{P} : \boldsymbol{\nabla} \boldsymbol{u} = 0$ from the vesicle membrane $\Gamma$ to the entire computational domain $\Omega$. 
Here, the surface gradient operator is defined as:
\begin{align} \label{eq:surface gradient operator}
    \boldsymbol{\nabla}_{\Gamma} = \boldsymbol{P} \boldsymbol{\nabla} = \boldsymbol{\nabla} - \boldsymbol{n} (\boldsymbol{n} \cdot \boldsymbol{\nabla}),
\end{align} 
where $\boldsymbol{P} = \boldsymbol{I} - \boldsymbol{n} \otimes \boldsymbol{n}$ denotes the projection tensor onto the tangential plane of the membrane, with the unit normal vector $\boldsymbol{n} = -\boldsymbol{\nabla} \phi / |\boldsymbol{\nabla} \phi|$ pointing from the interior to the exterior of the vesicle. The inextensibility constraint is enforced by introducing a local Lagrange multiplier $\lambda_{local}$ together with a harmonic regularization term, leading to \eqref{inextend_un}, where the parameter $\xi > 0$ controls the strength of the regularization. The regularized surface delta function is then defined as
\begin{align} \label{delta}
    \zeta(\phi) = \eta_\phi \left( \frac{f(\phi)}{\varepsilon_\phi} + \frac{\varepsilon_\phi}{2} |\nabla \phi|^2 \right),
\end{align}
where $f(\phi) = {(1 - \phi^2)^2}/{4}$ is a double-well potential, $\varepsilon_\phi$ is the membrane thickness, and $\eta_\phi$ is a scaling parameter that relates the diffuse interface to the sharp-interface model. The resulting constraint term contributes additional membrane tension forces in the momentum equation, completing the formulation of \eqref{ns_un}.

Furthermore, the velocity $\boldsymbol{u}$ is subject to Dirichlet boundary conditions, while the phase-field variable $\phi$ and its associated flux $\boldsymbol{q}_{\phi}$ satisfy no-flux boundary conditions:
	\begin{align}
\boldsymbol{u}=&\boldsymbol{u}_{\boldsymbol{bc}} ~~~~{\rm on}~~~~\partial\Omega,\\		\boldsymbol{n_b} \cdot \boldsymbol{\nabla} \phi=\boldsymbol{n_b} \cdot  \boldsymbol{q_{\phi}}=&0~~~~~~~~{\rm on}~~~~\partial\Omega,
	\end{align}
where $\boldsymbol{n_b}$ denotes the unit outward normal vector to the boundary $\partial\Omega$ of the rectangular computational domain $\Omega$.

\subsection{Governing equation for multicomponent on the vesicle membrane}	\

% We model the vesicle membrane $\Gamma$ as a two-component fluid composed of liquid-ordered ($l_o$) and liquid-ordered ($l_d$) phases with distinct physical properties, which is described by a second scalar phase-field variable $c_{\Gamma}(\boldsymbol{x}, t)$, where $c_{\Gamma} = 1$ represents the $l_d$ phase and $c_{\Gamma} = -1$ represents the $l_o$ phase. These two components are separated by a diffuse interface of thickness $\varepsilon_c$, across which $c_{\Gamma}$ transitions smoothly. The dynamics of phase separation are governed by a surface Cahn-Hilliard (CH) system, defined on $\Gamma$ \cite{lowengrub_numerical_2016}:

We model the vesicle membrane $\Gamma$ as a two-phase fluid consisting of liquid-ordered ($l_o$) and liquid-disordered ($l_d$) phases with distinct physical properties. The membrane composition is described by a second scalar phase-field variable $c_{\Gamma}$ defined on the membrane $\Gamma$, where $c_{\Gamma} = 1$ corresponds to the $l_d$ phase, and $c_{\Gamma} = -1$ to the $l_o$ phase. The two components are separated by a diffuse interface of thickness $\varepsilon_c$, across which $c_{\Gamma}$ transitions smoothly. The phase separation dynamics on $\Gamma$ are governed by the surface Cahn-Hilliard (CH) equation \cite{zhiliakov2021experimental}:
	\begin{align}  
		\label{ch_vs_sharp}
		 \left(c_{\Gamma}\right)_t + \boldsymbol{{\nabla}_{\Gamma}} \cdot  \left( c_{\Gamma} \boldsymbol{u} \right) &=  \boldsymbol{{\nabla}_{\Gamma}} \cdot \boldsymbol{q_c} ,
	\end{align}
where the diffusive flux of the membrane component, $\boldsymbol{q}_c$, will be specified later.

To couple the surface phase dynamics \eqref{ch_vs_sharp} with the bulk phase-field system \eqref{ns_un}-\eqref{inextend_un} in the rectangular domain $\Omega$, we employ the diffuse domain method \cite{teigen2009diffuse, li2009solving, ratz_pdes_2006}, which is an accurate and efficient numerical approach for solving partial differential equations (PDEs) on complex surfaces. The vesicle membrane $\Gamma$ is embedded into the larger, rectangular domain $\Omega$ using a regularized surface delta function $\zeta(\phi)$ derived from the phase-field variable $\phi$. This embedding eliminates the need for explicit interface tracking. 
Moreover, the method establishes a unified, energy-based framework by extending surface PDEs into the bulk domain $\Omega$ through the regularized delta function $\zeta(\phi)$ \eqref{delta}. Accordingly, the surface CH equation \eqref{ch_vs_sharp} is reformulated in $\Omega$ as:
\begin{align} 
		\label{ch_vs_un}
		 \left(\zeta(\phi) c_{\Gamma}\right)_t + \boldsymbol{\nabla} \cdot  \left(\zeta(\phi) c_{\Gamma} \boldsymbol{u}  \right) &=  \boldsymbol{\nabla} \cdot\left( \zeta(\phi) \boldsymbol{q_c} \right).
	\end{align}
 Details of this extension are provided in Appendix~\ref{appendix-ddm}. Notably, previous studies have demonstrated that in the diffuse domain method, the extended PDE \eqref{ch_vs_un} asymptotically converges to the original surface PDE \eqref{ch_vs_sharp} as the diffuse interface thickness approaches zero ($\varepsilon_{\phi}  \to 0$) \cite{li2009solving, yu2012extended, guo_diffuse_2021, guo_nonlinear_2023, lervag_analysis_2015, chadwick2018numerical, poulsen2018smoothed, yu2020higher}.

In addition, the membrane component $c_\Gamma$ and its flux $\boldsymbol{q_{c}}$ are subject to no-flux boundary conditions on the boundary of the rectangular domain $\Omega$:
\begin{align}\label{bc2}
\boldsymbol{n_b} \cdot  \boldsymbol{\nabla c_{\Gamma}} =\boldsymbol{n_b} \cdot \boldsymbol{q_{c}}=0~~~~{\rm on}~~~~\partial\Omega.
\end{align}  

Thanks to the diffuse domain method, the membrane component equation for $c_{\Gamma}$ \eqref{ch_vs_un} can be naturally coupled with the vesicle-fluid interaction system \eqref{ns_un}-\eqref{inextend_un} and solved simultaneously within the same  rectangular domain $\Omega$. The unknown terms $\boldsymbol{F}$, $\boldsymbol{q_{\phi}}$, and $\boldsymbol{q_{c}}$ in \eqref{ns_un}--\eqref{inextend_un} and \eqref{ch_vs_un} will be derived via an energy-variation approach based on the principle of free energy dissipation, thereby ensuring thermodynamic consistency.

\subsection{Free energy of the multicomponent vesicle-fluid system}\

The free energy $\mathcal{F}$ of a multicomponent vesicle immersed in a background fluid comprises contributions from both the fluid and the membrane. These include the kinetic energy of the background fluid $ \mathcal{F}_\text{K} $, the membrane bending energy $ \mathcal{F}_\text{B} $, the surface energy $ \mathcal{F}_\text{S} $, the line energy $ \mathcal{F}_\text{L} $, and a penalty term $ \mathcal{F}_\text{A} $ enforcing surface area conservation. The free energy is expressed as:
\begin{align} \label{eq:energy}
\mathcal{F} &= \mathcal{F}_\text{K} + \mathcal{F}_\text{B} + \mathcal{F}_\text{S} + \mathcal{F}_\text{L} + \mathcal{F}_\text{A} \\ \nonumber
&=  \int_{\Omega} \frac{1}{2} \rho |\boldsymbol{u}|^2 \, \mathrm{d}V + \int_{\Omega} \frac{1}{2 \varepsilon_\phi} \kappa_B(c_\Gamma) \omega_\phi^2 \, \mathrm{d}V + \int_{\Omega} \sigma_S(c_\Gamma) \zeta(\phi) \, \mathrm{d}V  \\ \nonumber
&+ \int_{\Omega} \sigma_L \zeta(c_\Gamma) \zeta(\phi) \, \mathrm{d}V \nonumber + \frac{M_s}{2} \frac{(S(\phi) - S(\phi_0))^2}{S(\phi_0)}. \nonumber
\end{align}

In the bending energy $\mathcal{F}_\text{B}$, the scalar field
\begin{align}
\omega_\phi = \eta_\phi \left( \frac{f'(\phi)}{\varepsilon_\phi} - \varepsilon_\phi \Delta \phi \right),
\end{align}
serves as a phase-field approximation of the membrane mean curvature. Here, zero spontaneous curvature is assumed and the contribution from Gaussian curvature is neglected. The bending rigidity $\kappa_B(c_\Gamma)$, associated with the bending energy $\mathcal{F}_\text{B}$, and the surface tension $\sigma_S(c_\Gamma)$, defined in the surface energy $\mathcal{F}_\text{S}$, are both modeled as functions of the membrane component field $c_\Gamma$:
\begin{align}
\kappa_B(c_\Gamma) = \frac{1 + c_\Gamma}{2} \kappa_{B_{l_d}} + \frac{1 - c_\Gamma}{2} \kappa_{B_{l_o}}, \quad 
\sigma_S(c_\Gamma) = \frac{1 + c_\Gamma}{2} \sigma_{S_{l_d}} + \frac{1 - c_\Gamma}{2} \sigma_{S_{l_o}}.
\end{align}
In the line energy $ \mathcal{F}_\text{L} $, the function
\begin{align}
 \zeta(c_\Gamma) = \eta_c \left( \frac{f(c_\Gamma)}{\varepsilon_c} + \frac{\varepsilon_c}{2} |\nabla c_\Gamma|^2 \right)   
\end{align}
serves as a regularized delta function that captures the diffuse interface between membrane components. Here, the double-well potential $ f(c_\Gamma) = (1 - c_\Gamma^2)^2/4 $ promotes phase separation, $\varepsilon_c$ represents the interfacial thickness between the components, $\eta_c$ is a scaling parameter, and $ \sigma_L $ denotes the line tension.
In the penalty energy $\mathcal{F}_\text{A}$, the vesicle surface area is defined as $S(\phi) = \int_\Omega \zeta(\phi) \, \mathrm{d}V$. The reference area $S(\phi_0)$ is measured at the initial time $t = 0$ and assumed constant during the simulation. Surface area conservation is enforced by this penalty term, with its strength controlled by the penalty coefficient $M_S$.

By applying an energy-variation approach, we derive explicit expressions for the terms $\boldsymbol{F}, \boldsymbol{q_\phi}$, and $\boldsymbol{q_c}$ that were previously left unspecified in the governing equations.  Detailed derivations are provided in Appendix~\ref{appendix-modle}.

\subsection{Nondimensionalization}\

To nondimensionalize the model, we choose the characteristic length, velocity, and time scales as $L$, $U$, and $L/U$, respectively.  Fluid properties are normalized with respect to the outer fluid (where $\phi = -1$), using characteristic viscosity $\mu_{{out}}$ and density $\rho_{{out}}$, while membrane properties are normalized by those of the $l_o$ phase ($c_\Gamma = -1$). The nondimensional form of the multicomponent vesicles-fluid model is then given by: 
\begin{align}\label{ns_1}
	(\rho \boldsymbol{u})_t +  \boldsymbol{\nabla} \cdot  (\rho\boldsymbol{u} \otimes \boldsymbol{u}) &= \frac{1}{Re}\boldsymbol{\nabla} \cdot (\mu \boldsymbol{D}(\boldsymbol{u})) -\boldsymbol{\nabla} p + \frac{1}{Re} \boldsymbol{F} +\boldsymbol{\nabla} \cdot\left(\zeta(\phi) \boldsymbol{P} \lambda_{{local}}\right)  , \\
	\label{ns_2}
	\boldsymbol{\nabla} \cdot \boldsymbol{u}&=0,\\
	\label{ch}
	\phi_t+\boldsymbol{\nabla} \cdot(\phi \boldsymbol{u} ) &= \frac{1}{Pe} \boldsymbol{\nabla} \cdot \boldsymbol{q_{\phi}},\\
	\label{inextend}
	\xi {\varepsilon_\phi}^{2} \boldsymbol{\nabla} \cdot\left(\phi^{2} \boldsymbol{\nabla} \lambda_{{local}}\right)+\zeta(\phi) \boldsymbol{P}: \boldsymbol{\nabla} \boldsymbol{u}&=0,\\
    	\label{ch_vs}
		 \left(\zeta(\phi) c_{\Gamma}\right)_t + \boldsymbol{\nabla} \cdot  \left(\zeta(\phi) c_{\Gamma} \boldsymbol{u}  \right)&= \frac{1}{Pe_{\Gamma}} \boldsymbol{\nabla} \cdot\left( \zeta(\phi) \boldsymbol{q_c}\right),
\end{align}
where	
\begin{align}   
		\boldsymbol{F}&=   g \boldsymbol{\nabla} \phi  - \zeta(\phi) c_{\Gamma} \boldsymbol{\nabla} \omega_c    , \\ 
        \boldsymbol{q_{\phi}} &=  m(\phi) \boldsymbol{\nabla} {g},\\
	\boldsymbol{q_c} &=  m(c_{\Gamma}) \boldsymbol{\nabla} {\omega_c},  
\end{align}
with
\begin{align}
	g&=     b\omega_{\phi}       
    +\frac{1}{Ca} \left({\kappa_B(c_{\Gamma})}{\eta_\phi} \frac{f^{\prime \prime}(\phi)}{{\varepsilon_\phi}^2}   \omega_{\phi}    -  {\eta_\phi} \Delta({\kappa_B(c_{\Gamma})}  \omega_{\phi} )   \right)    - {\eta_\phi} {\varepsilon_\phi} \boldsymbol{\nabla} b   \cdot \boldsymbol{\nabla} \phi    , \\ \nonumber
	b &= -\omega_{c} c_{\Gamma}  +Cn_{\Gamma} \sigma_L \delta(c_{\Gamma})  +  \mathcal{M}_{\text{s}} \frac{S(\phi)-S(\phi_0)}{S(\phi_0)} + Cs_{\Gamma} {\sigma_S(c_{\Gamma})} ,\\
	 \zeta(\phi)\omega_c	&=\frac{1}{Ca} \frac{{\varepsilon_\phi}^{-1}}{2}{\kappa_B}^{\prime}(c_{\Gamma}) {\omega_{\phi}}^{2}   + Cs {\sigma_S}^{\prime}(c_{\Gamma}) {\zeta(\phi)} \nonumber \\ 
	&\quad + Cn_{\Gamma} {\sigma_L}\zeta(\phi)\left({\eta_c}  \frac{f^{\prime}(c_{\Gamma})}{\varepsilon_c}-{\eta_c} {\varepsilon_c} \Delta c_{\Gamma}\right)- Cn_{\Gamma} {\sigma_L} {\eta_c} {\varepsilon_c} \boldsymbol{\nabla} \zeta(\phi) \cdot \boldsymbol{\nabla}{c_{\Gamma}}. \label{ov}
\end{align}

\eqref{ns_1} and \eqref{ns_2} describe the dynamics of the incompressible background fluid, while the evolution of the vesicle membrane is governed by \eqref{ch}. The local inextensibility of the vesicle is enforced via \eqref{inextend}, and the lateral phase separation of membrane components is governed by \eqref{ch_vs}.

The boundary conditions are imposed as follows: Dirichlet boundary conditions are applied to the fluid velocity $\boldsymbol{u}$, while no-flux boundary conditions are prescribed for the phase-field variables $\phi$ and $c_\Gamma$, together with their respective fluxes $\boldsymbol{q}_\phi$ and $\boldsymbol{q}_c$:
\begin{align}\label{bcu}
\boldsymbol{u}=&\boldsymbol{u}_{\boldsymbol{bc}} ~~~~{\rm on}~~~~\partial\Omega,\\ \label{bcphi}
\boldsymbol{n_b} \cdot \boldsymbol{\nabla} \phi=\boldsymbol{n_b} \cdot  \boldsymbol{q_{\phi}}=&0~~~~~~~~{\rm on}~~~~\partial\Omega, \\ \label{bcc}
\boldsymbol{n_b} \cdot  \boldsymbol{\nabla c_{\Gamma}} =\boldsymbol{n_b} \cdot \boldsymbol{q_{c}}=&0~~~~~~~~{\rm on}~~~~\partial\Omega,
	\end{align}
where $\boldsymbol{n_b}$ denotes the unit outward normal vector to the boundary $\partial\Omega$ of the rectangular computational domain $\Omega$.

The vesicle dynamics described by \eqref{ns_1}--\eqref{ov} are governed by a set of dimensionless parameters that characterize the coupling between the background fluid and membrane dynamics, the lateral phase separation, and the influence of line tension and component-dependent physical properties. These dimensionless parameters are summarized in table~\ref{tab:parameters}.

\begin{table}
    \centering
    \vspace{0.5em}
    \begin{tabular}{lll}
        \toprule
        \textbf{Symbol} & \textbf{Expression} & \textbf{Description} \\
        \midrule
        & & \textit{Fluid Parameters} \\
        $\rho$ & -- & fluid density, with $\rho_{{in}} = \rho_{{out}}$ \\
        $\mu(\phi)$ & $(1+\phi)\mu_{{in}}/2+(1-\phi)\mu_{{out}}/2$ & fluid viscosity \\
        \midrule
        & &  \textit{Diffuse interface membrane parameters} \\
        $\varepsilon_{\phi}$ & -- & diffuse interface thickness of the membrane \\
        $m(\phi)$ & $D_\phi\left(1 - {\phi}^2\right)^2 $ &   mobility of the phase-field $\phi$ (diffusion coefficient $D_\phi$) \\
        $M_S$ & -- & penalty coefficient enforcing surface area conservation \\
        $\xi$ & -- & relaxation parameter for inextensibility \\
        \midrule
        & & \textit{Membrane component parameters} \\
        $\varepsilon_c$ & -- & diffuse interface thickness of membrane component \\
        $m(c_{\Gamma})$ & $D_c\left(1 - {c_{\Gamma}}^2\right)^2 $ & mobility of membrane component $c_{\Gamma}$ (diffusion coefficient $D_c$) \\
        $\kappa_B(c_{\Gamma})$ & $({1 + c_{\Gamma}})\kappa_{B_{l_d}}/{2}  + ({1 - c_{\Gamma}})\kappa_{B_{l_o}}/{2}$ & component-dependent bending rigidity \\
        $\sigma_S(c_{\Gamma})$ & $({1 + c_{\Gamma}})\sigma_{S_{l_d}}/{2}  + ({1 - c_{\Gamma}})\sigma_{S_{l_o}}/{2} $ & component-dependent surface tension \\
        $\sigma_L$ & -- & line tension between components\\
        \midrule
        & & \textit{Dimensionless Parameters} \\
        $Re$ & ${\rho_{{out}} U L}/{\mu_{{out}}}$ & Reynolds number \\
        $Ca$ & ${\mu_{{out}} U L^2}/{\kappa_{B_{l_o}}}$ & bending capillary number \\
        $Pe$ & ${L^2}/{D_\phi \mu_{{out}}}$ & P\'eclet number for $\phi$ \\
        $Pe_{\Gamma}$ & ${L}/{D_c \mu_{{out}}}$ & P\'eclet number for $c_{\Gamma}$ \\
        $Cn_{\Gamma}$ & ${\sigma_L}/{\mu_{{out}} U L}$ & line tension to viscous force ratio \\
        $Cs_{\Gamma}$ & ${\sigma_{S_{l_o}}}/{\mu_{{out}} U}$ & surface tension to viscous force ratio \\
        $\kappa_B$ & $\kappa_{B_{l_d}} / \kappa_{B_{l_o}}$ & bending rigidity contrast \\
        $\sigma_S$ & $\sigma_{S_{l_d}} / \sigma_{S_{l_o}}$ & surface tension contrast \\
        $\mathcal{M}_{\text{s}}$ & $M_S / (\mu_{{out}} U)$ & dimensionless surface area penalty coefficient \\
        $v$ & ${3V}/{4\pi ({A}/{4\pi})^{3/2}}$ & reduced volume ($v = 1$ for sphere, $v < 1$ for deflated shape) \\
        $\gamma$ & $\dot{\gamma}L/U$ & dimensionless shear rate \\
        \midrule
        & & \textit{Geometric parameters} \\
        $\theta$ & $\theta \in [-\pi/2, \pi/2]$ & inclination angle between vesicle’s long axis and flow direction \\
        \bottomrule
    \end{tabular}
     \caption{Physical and dimensionless parameters used in the simulations.}
    \label{tab:parameters}
\end{table}

Our model integrates key physical features, including membrane component-dependent bending rigidity heterogeneity, surface tension heterogeneity, line tension, inextensibility, and viscosity contrasts between the internal and external fluids. Its thermodynamically consistent formulation ensures free energy dissipation and physical consistency, enabling the simulation of complex dynamic processes under realistic biological conditions. This comprehensive framework provides a powerful tool for investigating vesicle behavior under shear flow and for elucidating the role of membrane heterogeneity in governing nonequilibrium dynamics.

\section{Numerical Validations} \label{sec.3 validation}\

To validate the model, we conducted systematic comparisons between our simulations and two sets of experimental observations:  (i) phase separation dynamics on giant unilamellar vesicles (GUVs) reported in \cite{wang2022lipid}, and (ii) dynamic behaviors of pre-positioned multicomponent vesicles under shear flow reported in \cite{tusch_when_2018}. The numerical method is detailed in Appendix~\ref{appendix-numerical method}. At the fully discrete level, we employ BSAM -- a highly efficient block-structured adaptive Full Approximation Scheme (FAS) multigrid solver -- for the simulations. The solver combines a staggered finite-difference method with a FAS multigrid algorithm and is specifically designed to address diffuse-interface problems \cite{guo2017mass}. Within this framework, the velocity components $\boldsymbol{u} = (u, v, w)$ are stored at edge centers, while all scalar quantities, including pressure and phase-field variables, are stored at cell centers. Details of the spatial discretization can be found in \cite{guo2017mass}. Our simulations exhibit excellent qualitative and quantitative agreement with both experiments. These validations confirm the accuracy, effectiveness, and predictive capability of the proposed model in capturing phase behaviors, vesicle-fluid coupling, and dynamic transitions of multicomponent vesicles under shear flow.

% The numerical method is implemented by using BSAM, an accurate and efficient finite difference FAS multigrid solver mainly designed for solving diffuse interface problems.

\subsection{ Phase separation on stationary multicomponent vesicles}\label{sec.3.1 phase separation}\

We first investigate phase separation dynamics on the membrane of multicomponent vesicles in comparison with the experimental results reported in \cite{wang2022lipid}. That study examined the phase behavior of GUVs and provided detailed observations of domain evolution. To facilitate a direct comparison, we simplify our setup by assuming a stationary vesicle with a fixed spherical shape, focusing on capturing the essential features of membrane phase separation.

% Our simulations reproduce the experimentally observed phase separation dynamics, confirming the validity and effectiveness of the model in describing multicomponent vesicle phase behavior.

The computational domain is defined as $\bar{\Omega} = [-1.5, 1.5]^3$ and discretised using a uniform Cartesian grid of $128^3$. The fixed spherical vesicle is represented by a phase-field with radius $r_0=1$, corresponding to a physical vesicle radius $R = 10 \,\mu$m \cite{wang2022lipid}, and membrane thickness $\varepsilon_\phi = 0.02$. The initial phase-field variable $\phi(\boldsymbol{x}, 0)$ is prescribed as
\begin{align}\label{initial shape}
\phi(\boldsymbol{x}, 0) = \tanh\left(\frac{r_0 - r}{2\sqrt{2}\varepsilon_\phi}\right), \quad
r = \sqrt{x^2 + y^2 + z^2}, \quad \boldsymbol{x} =(x,y,z) \in \bar{\Omega},
\end{align}
representing a smooth diffuse interface centered around the vesicle surface. To model an initially homogeneous liposome, we first generate random numbers $ c_{\text{rand}}(\boldsymbol{x})$ with a uniform distribution between 0 and 1. Then, we apply the following cutoff for the initial surface fraction $ c_\Gamma(\boldsymbol{x}, 0) $:
\begin{align} \label{ic-phase-sep-c_gamma}
    c_{\Gamma}(\boldsymbol{x}, 0) = 
    \begin{cases}
        ~~1 \quad (l_{d}), & \text{if } c_{\text{rand}}(\boldsymbol{x}) \in [0, a_{l_d}], \\
        -1 \quad (l_{o}), & \text{if } c_{\text{rand}}(\boldsymbol{x}) \in (a_{l_d}, 1],
    \end{cases}
\end{align}
where $a_{l_d}$ and $a_{l_o} = 1 - a_{l_d}$ represent the prescribed area fractions of the $l_d$ and $l_o$ phases, respectively.

No-flux boundary conditions are applied to the phase-field variables $\phi$, $c_{\Gamma}$ and $\omega_c$ at the domain boundary $\partial \Omega$.

The model parameters are chosen to match those reported in \cite{wang2022lipid}: line tension $\sigma_L = 1.2 \,\text{pN}$ and diffusion coefficient $D_c = 10^{-5}\,\text{cm}^2/\text{s}$. The lipid mobility is adjusted such that $Cn_\Gamma/Pe_\Gamma = 0.12$. Bending rigidity and surface tension are assumed to be uniform across the $l_o$ and $l_d$ phases, with $\kappa_B = \sigma_S = 1$. The interfacial thickness of membrane components is set to $\varepsilon_c = 0.02$, fitted to match experimental GUVs.

Two area fractions, $a_{l_d} = 0.3$ and $a_{l_d} = 0.7$, are considered to match the experimental conditions in \cite{wang2022lipid}. To account for stochastic variability, we perform 10 independent simulations with different random seeds of $c_{\Gamma}$ for each area fraction.

\begin{figure}[h]
   \centering
   \includegraphics[width=0.8\linewidth]{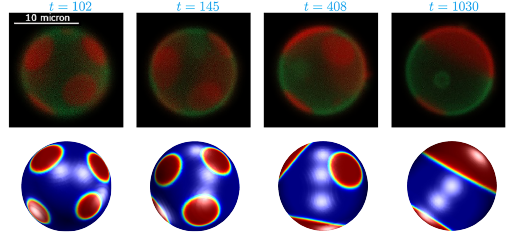}
   \caption{Phase separation on GUVs at $a_{l_d} = 0.3$: experimental snapshots (top) from \cite{wang2022lipid} and numerical results (bottom), showing $l_o$ domains (blue) forming within the $l_d$ phase (red).
   }
   \label{fig:73od}
\end{figure}

\begin{figure}[h]
   \centering
   \includegraphics[width=0.8\linewidth]{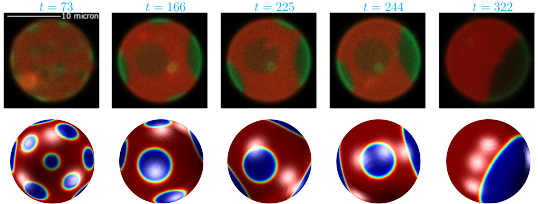}
   \caption{Phase separation on GUVs at $a_{l_d} = 0.7$: experimental snapshots (top) from \cite{wang2022lipid} and numerical results (bottom), showing $l_d$ domains (red) forming within the $l_o$ phase (blue).
   }
   \label{fig:37od}
\end{figure}

Figures~\ref{fig:73od} and \ref{fig:37od} show the phase separation process on the membrane for area fractions $ a_{l_d} = 0.3 $ and $ a_{l_d} = 0.7 $, respectively, compared with the experimental results from \cite{wang2022lipid}. During phase separation, membrane components coarsen, forming distinct, stable domains, as illustrated in the far-right images of both figures~\ref{fig:73od} and \ref{fig:37od} that closely match the experimental snapshots.

\begin{figure}[h]
   \centering
   \includegraphics[width=0.8\linewidth]{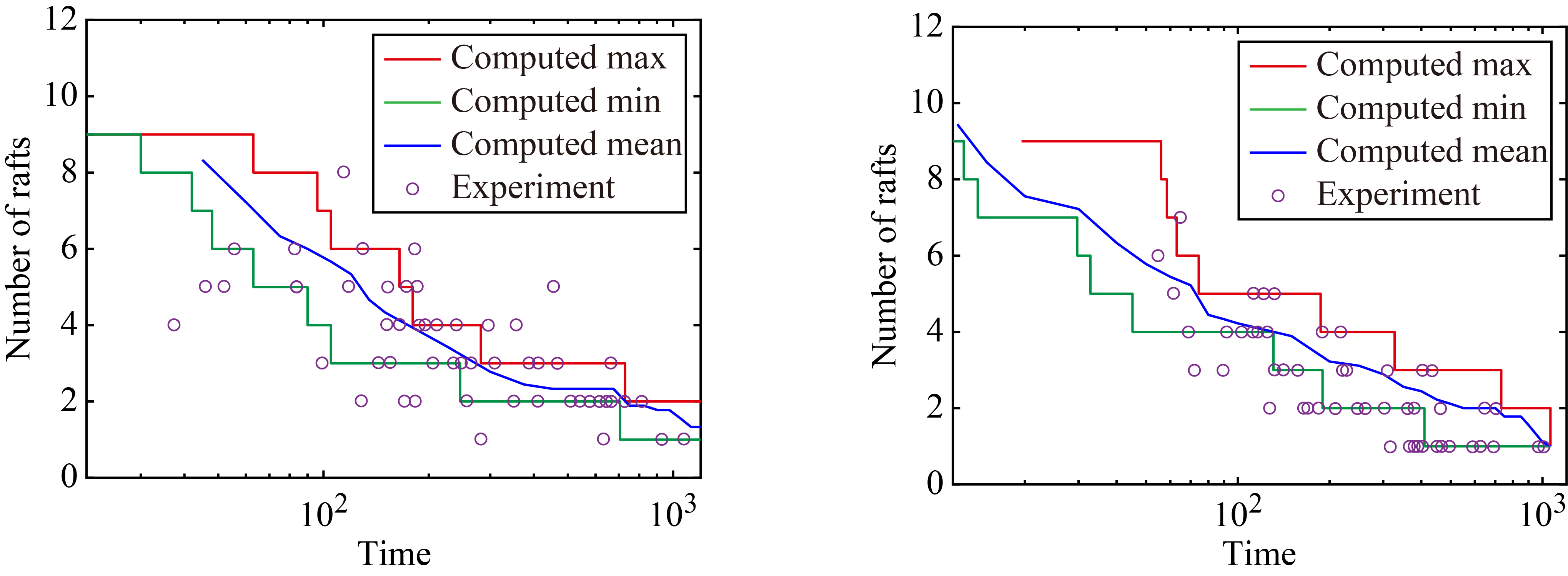}
   \caption{Time evolution of number of lipid rafts for $a_{l_d} = 0.3$ (left) and $a_{l_d} = 0.7$ (right). Markers represent experimental data; solid lines represent numerical results (blue: average, green: minimum, red: maximum).}
   \label{fig:phase-num}
\end{figure}

To further validate the accuracy of our model, we compare the time evolution of the number of $l_d$ domains predicted by our numerical simulations with the experimental measurements reported in \cite{wang2022lipid}. As shown in figure~\ref{fig:phase-num}, for both area fractions under investigation ($a_{l_d} = 0.3$ and $a_{l_d} = 0.7$), the experimental data (markers) are generally bounded by the minimum (green) and maximum (red) values from the ensemble of our simulations. The average numerical result (blue) also lies in close agreement with the experimental trend.

% These comparative results, in terms of both domain number and total perimeter evolution, demonstrate the robustness and predictive capability of our model in accurately capturing the lipid phase separation dynamics.

\begin{figure}[h]
   \centering
   \includegraphics[width=0.8\linewidth]{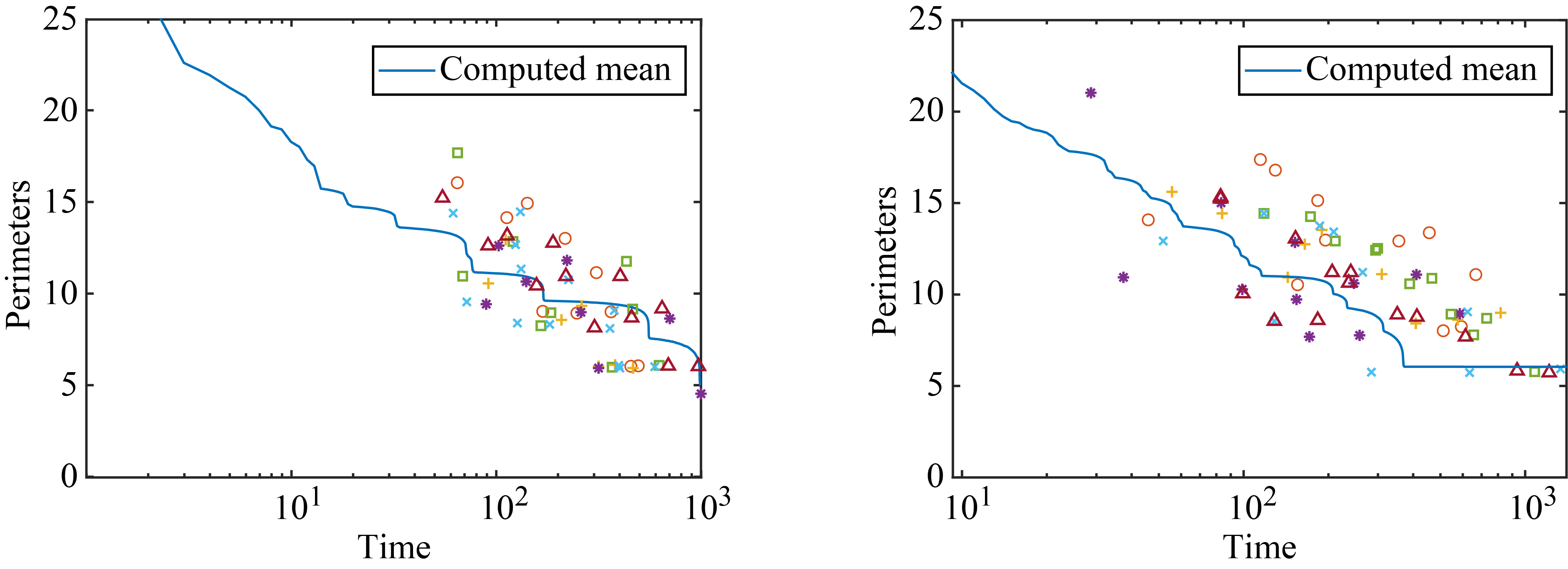}
   \caption{Time evolution of the total perimeter of lipid domains at $a_{l_d} = 0.3$ (left) and $a_{l_d} = 0.7$ (right). Markers represent experimental data; solid lines represent numerical results.}
   \label{fig.perimeters}
\end{figure}

In addition, we further compare the evolution of the total perimeter of lipid domains, quantified by $\int_\Omega \zeta(\phi) \zeta(c_\Gamma)\, \mathrm{d}V$ in our simulations, with the experimental data, as shown in figure~\ref{fig.perimeters}. The simulation results (solid lines), averaged over all numerical simulations, exhibit excellent agreement with the experimental measurements  (markers) for both $a_{l_o} = 0.7$ and $a_{l_o} = 0.3$.

% This strong consistency of both domain number and total perimeter evolution across the entire time course indicates that our model can reliably capture the domain coarsening dynamics observed in experiments.

\begin{figure}[h]
   \centering
   \includegraphics[width=0.8\linewidth]{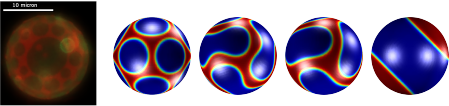}
   \caption{Experimental images (left) from \cite{wang2022lipid} and numerical results (right) showing the time evolution of ${l_o}$ domains (blue) within the ${l_d}$ phase (red). The simulation is initialized with 12 symmetrically placed ${l_o}$ domains and an area fraction $a_{l_o} = 0.64$.}
   \label{fig:73od-12}
\end{figure}

\begin{figure}[h]
   \centering
   \includegraphics[width=0.8\linewidth]{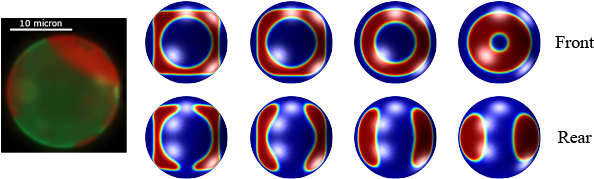}
   \caption{Experimental images (left) from \cite{wang2022lipid} and numerical results (right) with front and rear views over time, showing a combination of ${l_o}$ phase (blue) and ${l_d}$ phase (red). The simulation is initialized with 6 symmetrically placed $l_o$ domains and an area fraction $a_{l_o} = 0.65$.}
   \label{fig:73od-6}
\end{figure}

As reported in \cite{wang2022lipid}, three distinct structural patterns were observed experimentally at an area fraction of $a_{l_d} = 0.3$ prior to equilibrium: (i) the minority $l_d$ phase domains (red) embedded within the majority $l_o$ phase (blue), as shown in figure~\ref{fig:73od}; (ii) majority $l_o$ phase domains (blue) enclosed within the minority $l_d$ phase (red), as illustrated in figure~\ref{fig:73od-12}; and (iii) a mixed pattern featuring both types of domains, as depicted in figure~\ref{fig:73od-6}.

Patterns (ii) and (iii) were attributed to the persistence of a metastable third phase, which influences domain coarsening and leads to fluctuations in domain area fractions before equilibrium \cite{van1984ostwald,
anderson2002insights}. For instance, the measured $l_o$ area fraction in case (ii) was about 6\% lower than its equilibrium value.

Following \cite{wang2022lipid}, we reproduce the experimentally observed patterns by adjusting the initial conditions: (i) prescribing pre-positioned, symmetrically arranged $l_o$ domains within an $l_d$ background, and (ii) reducing the $l_o$ area fraction by approximately $6\%$. Accordingly, the second pattern--$l_o$ domains fully enclosed within the $l_d$ phase (figure~\ref{fig:73od-12})--is reproduced by initializing 12 symmetrically placed $l_o$ domains with an area fraction $a_{l_o}=0.64$. The third pattern, featuring coexisting $l_o$ and $l_d$ domains (figure~\ref{fig:73od-6}), is obtained by initializing 6 symmetrically distributed $l_o$ domains and setting $a_{l_o}=0.65$.

% This strong consistency of both domain number and total perimeter evolution across the entire time course indicates that our model can reliably capture the domain coarsening dynamics observed in experiments.

The quantitative and qualitative agreement between our reproduced simulations and the experimental observations demonstrates the effectiveness and accuracy of our model in capturing the phase dynamics associated on the vesicle membrane.

\subsection{Dynamics of two-component vesicle  under shear flow}\label{sec.3.2 dynamics-exp}\

To assess the capability of the model in capturing membrane--fluid interactions, we reproduce the experimental observations of multicomponent GUVs under shear flow reported in \cite{tusch_when_2018}. These experiments revealed dynamic behaviours such as tank treading, swinging, and tumbling, driven by asymmetric phase distributions and periodic elastic energy storage along the contact line.

\begin{figure}[h]
   \centering
   \includegraphics[width=0.4\linewidth]{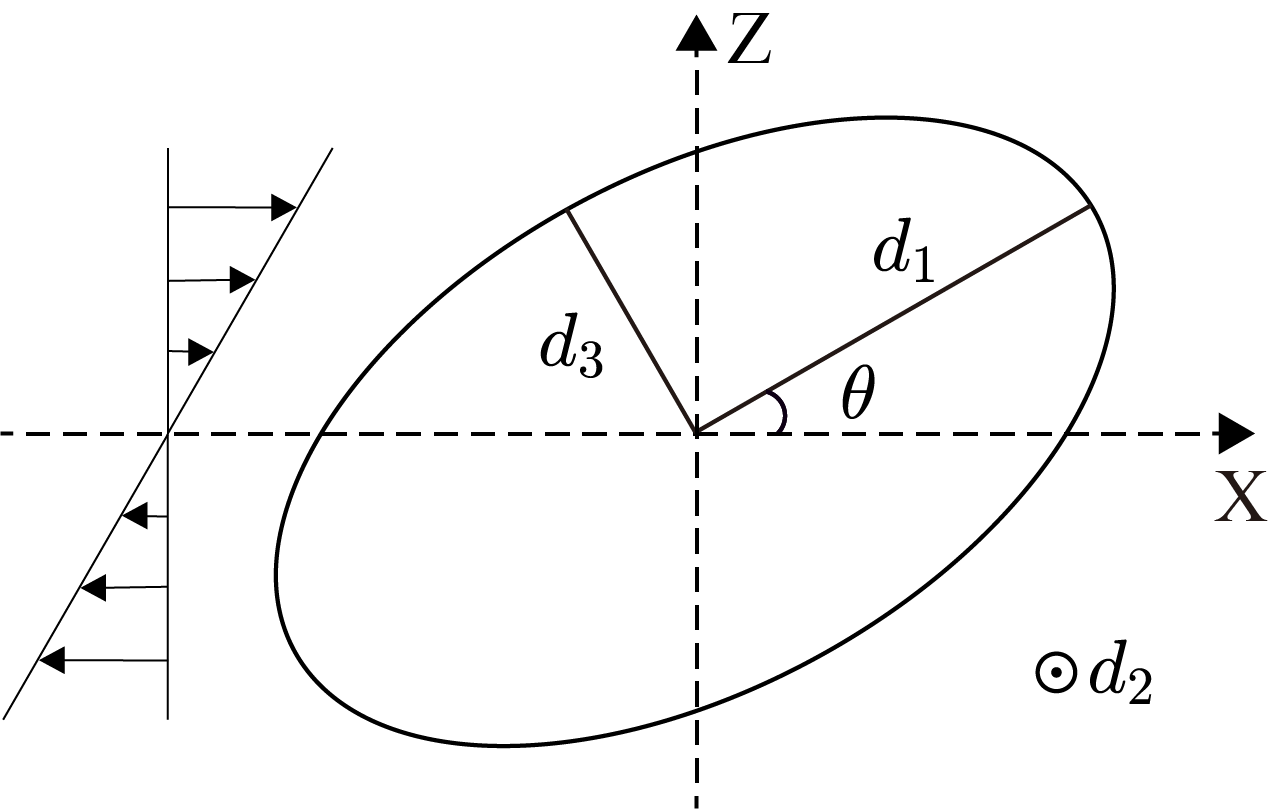}
   \caption{Geometric parameters used to characterize vesicle dynamics under shear flow: the principal axes of the vesicle ($d_1$, $d_2$, $d_3$), and the inclination angle $\theta$ defined as the angle between the long axis $d_1$ and the flow direction (x-axis).}
   \label{fig:angle-axis}
\end{figure}

In our simulations, the computational domain is defined as $\bar{\Omega} = [-2, 2]^3$, discretized on a uniform grid with $128^3$. To match the experimental setup, the initial vesicle shape is modeled as a prolate ellipsoid with a specified reduced volume. The initial phase variable is given by
\begin{align}\label{prolate shape}
\phi(\boldsymbol{x}, 0) = \tanh\left(\frac{r_0 - r}{2\sqrt{2}\varepsilon_\phi}\right), \quad
r = \frac{x^2}{d_1^2} + \frac{y^2}{d_2^2} + \frac{z^2}{d_3^2}, \quad \boldsymbol{x} = (x, y, z) \in \bar{\Omega},
\end{align}
where $d_1 > d_2 = d_3$, resulting in elongation along the $x$-axis (see figure~\ref{fig:angle-axis}). The membrane component $c_{\Gamma}$ is initially pre-positioned at the right-hand tip of the vesicle along the $x$-axis, with a prescribed area fraction $a_{l_d}$ representing the fraction of the $l_d$ phase. 

% A linear shear flow $\boldsymbol{u}(\boldsymbol{x}, 0) = (\gamma z, 0, 0)$ is imposed, with $\gamma$ as the shear rate. The flow is driven via wall Dirichlet boundary conditions: $\boldsymbol{u}_{bc} = (\gamma z, 0, 0)$ applied to the top and bottom boundaries in the $z$-direction. No-flux boundary conditions are enforced for all phase-field variables ($\phi$, $\omega_\phi$, $c_\Gamma$, $\omega_c$) in the $z$-direction, while periodic boundary conditions are applied in the $x$- and $y$-directions for all variables.

The shear flow is introduced by imposing boundary conditions. Specifically, Dirichlet boundary conditions $\boldsymbol{u}_{bc} = (\gamma z, 0, 0)$ are imposed at the top and bottom boundaries in the $z$-direction, where $\gamma$ denotes the dimensionless shear rate (see table~\ref{tab:parameters}). No-flux boundary conditions are enforced for all phase-field variables ($\phi$, $\omega_\phi$, $c_\Gamma$, $\omega_c$) in the $z$-direction, while periodic boundary conditions are applied in the $x$- and $y$-directions for all variables.

Physical parameters are mainly taken from \cite{tusch_when_2018}: mean vesicle radius $R = 10 \,\mu\text{m}$, fluid viscosity $\mu_{in} = \mu_{out} = 1.2 \times 10^{-3} \,\text{Pa}\cdot\text{s}$, bending rigidity $\kappa_{B_{l_o}} = \kappa_{B_{l_d}} = 10^{-19}\,\text{J}$, and shear rate $\dot{\gamma} = 3\,\text{s}^{-1}$. Fluid densities are taken as $\rho_{in} = \rho_{out} = 1000 \,\text{kg}/\text{m}^3$ \cite{luby1999cytoarchitecture}. Surface tensions are fixed at $\sigma_{S_{l_d}} = \sigma_{S_{l_o}} = 10^{-6}\,\text{N}/\text{m}$ to ensure uniform membrane properties. The line tension $\sigma_L$ is varied as $0.1,\,1,$ and $4\,\text{pN}$ to reproduce the three dynamical regimes observed experimentally.

\begin{figure}[h]
   \centering
 \includegraphics[width=0.8\linewidth]{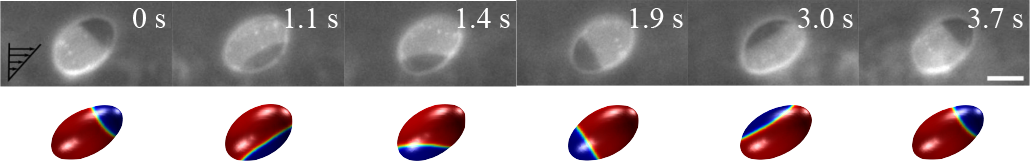}
   \caption{Experimental images (top) from \cite{tusch_when_2018} and numerical results (bottom) showing tank treading and swinging at \( Cn_\Gamma = 0.2 \), for a vesicle with reduced volume $v = 0.95$, flattening $f = 0.3$, and area fraction $a_{l_d} = 0.7$.}  
   \label{fig:c-0.7}
\end{figure}

\begin{figure}[h]
   \centering
   \includegraphics[width=0.8\linewidth]{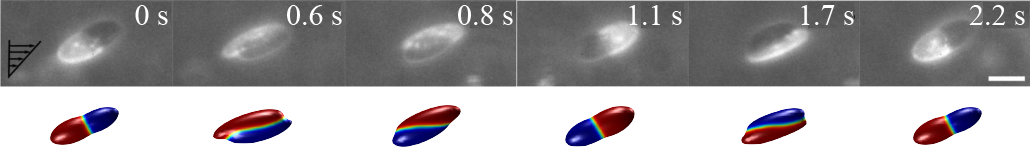}
   \caption{Experimental images (top) from \cite{tusch_when_2018} and numerical results (bottom) showing strong deformation during tank treading and swinging at \( Cn_\Gamma = 2 \), for a vesicle with reduced volume \( v = 0.8 \), flattening parameter \( f = 0.65 \), and area fraction \( a_{l_d} = 0.6 \).}
      \label{fig:c-0.35}
\end{figure}

\begin{figure}[h]
   \centering
   \includegraphics[width=0.8\linewidth]{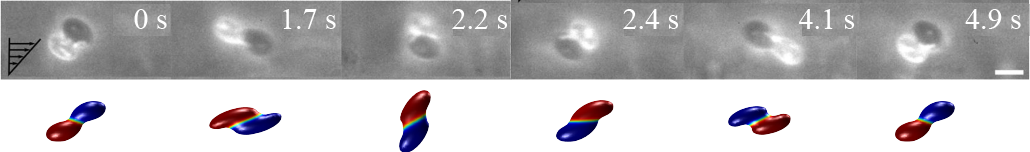}
   \caption{Experimental images (top) from \cite{tusch_when_2018} and numerical results (bottom) showing the tumbling at $Cn_\Gamma=8$, for a vesicle with reduced volume \( v = 0.7 \), flattening parameter \( f = 0.75 \), and area fraction \( a_{l_d} = 0.5 \). }
   \label{fig:c-0.25}
\end{figure}

Using $R$ as the characteristic length and $\dot{\gamma}R$ as the velocity scale, the resulting dimensionless parameters are: Reynolds number $Re = 3 \times 10^{-4}$, bending capillary number $Ca = 36$, P\'eclet numbers $Pe = 1000$ and $Pe_\Gamma = 10$, dimensionless surface tension strength $Cs = 27$, and surface area penalty coefficient $\mathcal{M}_\mathrm{S} = 2000$. The Cahn number is set as $Cn_\Gamma = 0.2,\,2,\,8$ for the three regimes (figures~\ref{fig:c-0.7}--\ref{fig:c-0.25}). Interfacial thickness parameters $\varepsilon_c = 0.05$ for membrane components and $\varepsilon_\phi = 0.05$ for the membrane are fitted to experimental GUVs.

Figures~\ref{fig:c-0.7}--\ref{fig:c-0.25} illustrate vesicle dynamics under different reduced volumes, showing close agreement with the experimental results from \cite{tusch_when_2018}. In figure~\ref{fig:c-0.7}, the initial ellipsoid with \( d_1 = 1 \), \( d_2 = d_3 = 0.7 \), yields \( v \approx 0.95 \), flattening parameter \( f = ({d_1 - d_2})/{d_1} = 0.3 \), and area fraction \( a_{l_d} = 0.7 \). At $Cn_\Gamma=0.2$, the vesicle exhibits tank treading and swinging characterized by a steady prolate shape with oscillatory inclination angle.

At a moderate reduced volume \( v \approx 0.8 \), shown in figure~\ref{fig:c-0.35}, with \( d_1=1, d_2 = d_3 = 0.35 \), \( f = 0.65 \), and \( a_{l_d} = 0.6 \), the vesicle exhibits pronounced deformation at $Cn_\Gamma=2$. It appears elliptical when the \( l_o \)-\( l_d \) interface aligns with the short axis and resembles two adhered vesicles when aligned with the long axis, exhibiting tank treading and swinging  accompanied by large deformations.

At lower volume $v \approx 0.7$ (figure~\ref{fig:c-0.25}), with $d_1=1, d_2 = d_3 = 0.25$, $f = 0.75$, and $a_{l_d} = 0.5$, the vesicle transitions to a tumbling regime at $Cn_\Gamma=8$. The interface length between $l_o$ and $l_d$ is markedly reduced along the long axis due to line tension, which drives the system to minimize line energy, facilitating large deformations and promoting tumbling under shear.

\begin{figure}[h]
   \centering
   \includegraphics[width=0.5\linewidth]{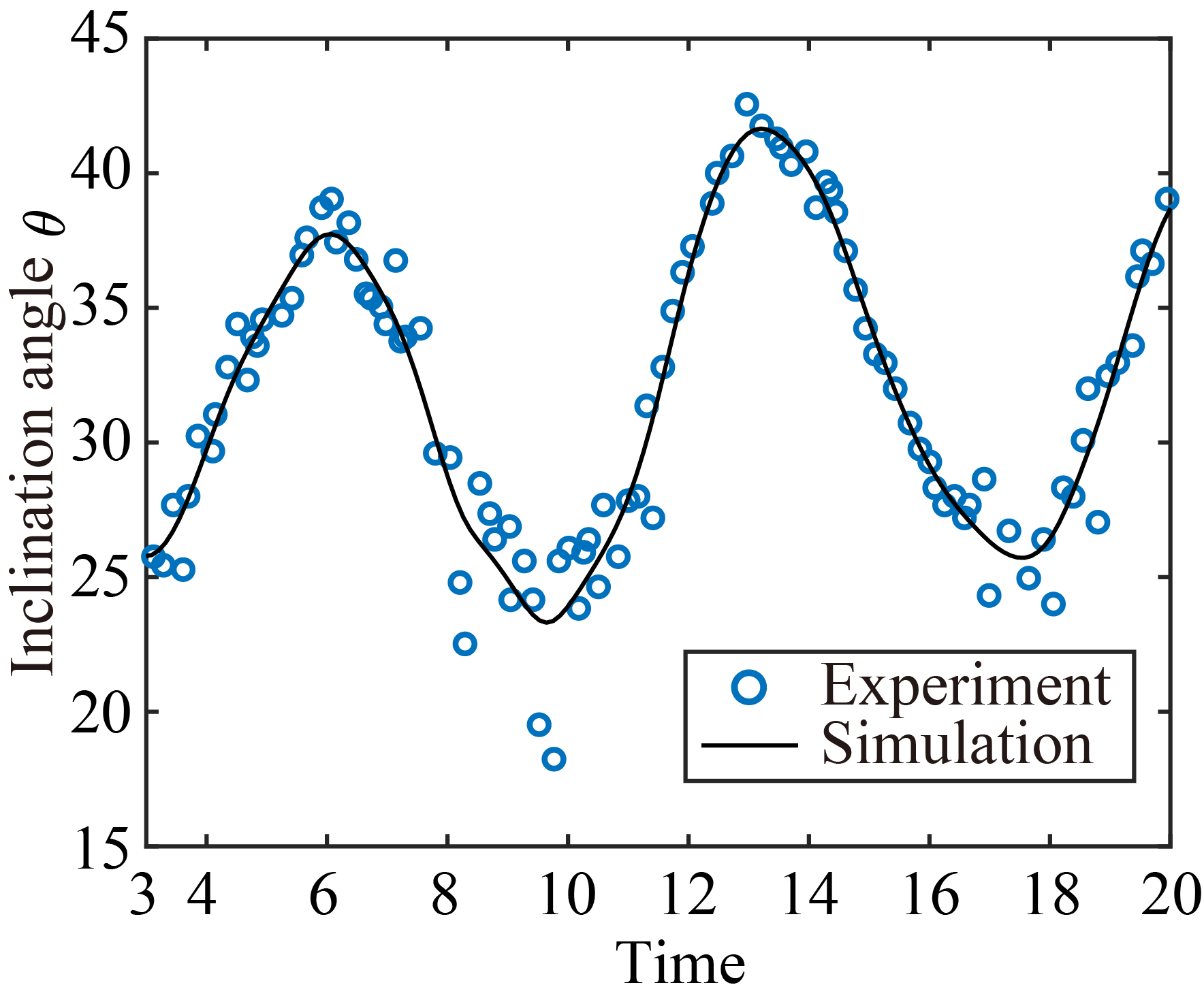}
   \caption{Time evolution of inclination angle $ \theta $ for a vesicle with reduced volume $ v = 0.95 $, flattening parameter $ f = 0.3 $, and area fraction $ a_{l_d} = 0.7 $ at $Cn_\Gamma=0.2$, comparing numerical results (solid line) with experimental data (circles) from \cite{tusch_when_2018}.}
   \label{fig:angle_exp}
\end{figure}

To further validate our model, we track the time evolution of the inclination angle $\theta$ of a vesicle undergoing tank treading and swinging behaviors at reduced volume $v = 0.95$ and area fraction $a_{l_d} = 0.7$ (see figure~\ref{fig:c-0.7}). The inclination angle $\theta$ is defined as the angle between the vesicle’s long axis and the flow direction, constrained within $\theta \in \left[-{\pi}/{2}, {\pi}/{2}\right]$ (see figure~\ref{fig:angle-axis}).
As shown in figure~\ref{fig:angle_exp}, the simulated results (solid line) closely match the experimental measurements (circles) reported in \cite{tusch_when_2018}, capturing the characteristic oscillations of the tank treading and swinging regime with a mean angle of approximately $30^\circ$. The numerical results remain consistently within the experimental variability, thereby confirming not only the accuracy and robustness of the model but also its predictive capability in reproducing flow-induced vesicle dynamics.

\section{Numerical experiments}\label{sec.4 simulations} \

In this section, we investigate the flow-induced dynamics of multicomponent vesicles, with a particular focus on the coupling between membrane heterogeneity and hydrodynamic forces. Through comprehensive numerical simulations, we demonstrate that the competition among shear flow, bending rigidity heterogeneity, and surface tension heterogeneity governs complex vesicle behaviors, including shape deformation, domain evolution, and transitions between distinct dynamic regimes.

\subsection{Bending heterogeneity driven dynamics}\label{sec.4.1:kb-ca}\

We systematically explore the coupling between shear flow and bending heterogeneity in multicomponent vesicles by varying the bending rigidity contrast $\kappa_B = \kappa_{B_{l_d}} / \kappa_{B_{l_o}}$ and the bending capillary number $Ca = \mu_{{out}} U L^2 / \kappa_{B_{l_o}}$, both readily tunable in experiments. 

The numerical setup closely follows Section~\ref{sec.3.2 dynamics-exp}. The vesicle is initialized as a prolate ellipsoid with principal axes $d_1 = 1$ and $d_2 = d_3 = 0.565$, as defined in \eqref{prolate shape}, corresponding to a reduced volume $v \approx 0.91$. Two symmetric $l_d$-phase domains (red) are pre-positioned at the vesicle tips along the $x$-axis, with an initial area fraction $a_{l_d} = 0.4$ for the membrane component field $c_{\Gamma}$. The dimensionless shear rate is set to $\gamma = 1$.

Moreover, the dimensionless parameters are specified as follows: Reynolds number $Re = 10^{-3}$, Cahn number $Cn_{\Gamma} = 0.05$, P\'eclet numbers $Pe = 1000$ and $Pe_\Gamma = 1 $, dimensionless surface-tension strength $Cs_{\Gamma} = 10$, surface-area penalty coefficient $\mathcal{M}_\mathrm{S} = 2000$, and surface tension contrast $\sigma_S = 1$. In all cases, the $l_d$ phase is assumed to be softer than the $l_o$ phase, i.e., $\kappa_{B_{l_d}} < \kappa_{B_{l_o}}$.

\begin{figure}[h]
   \centering
   \includegraphics[width=0.8\linewidth]{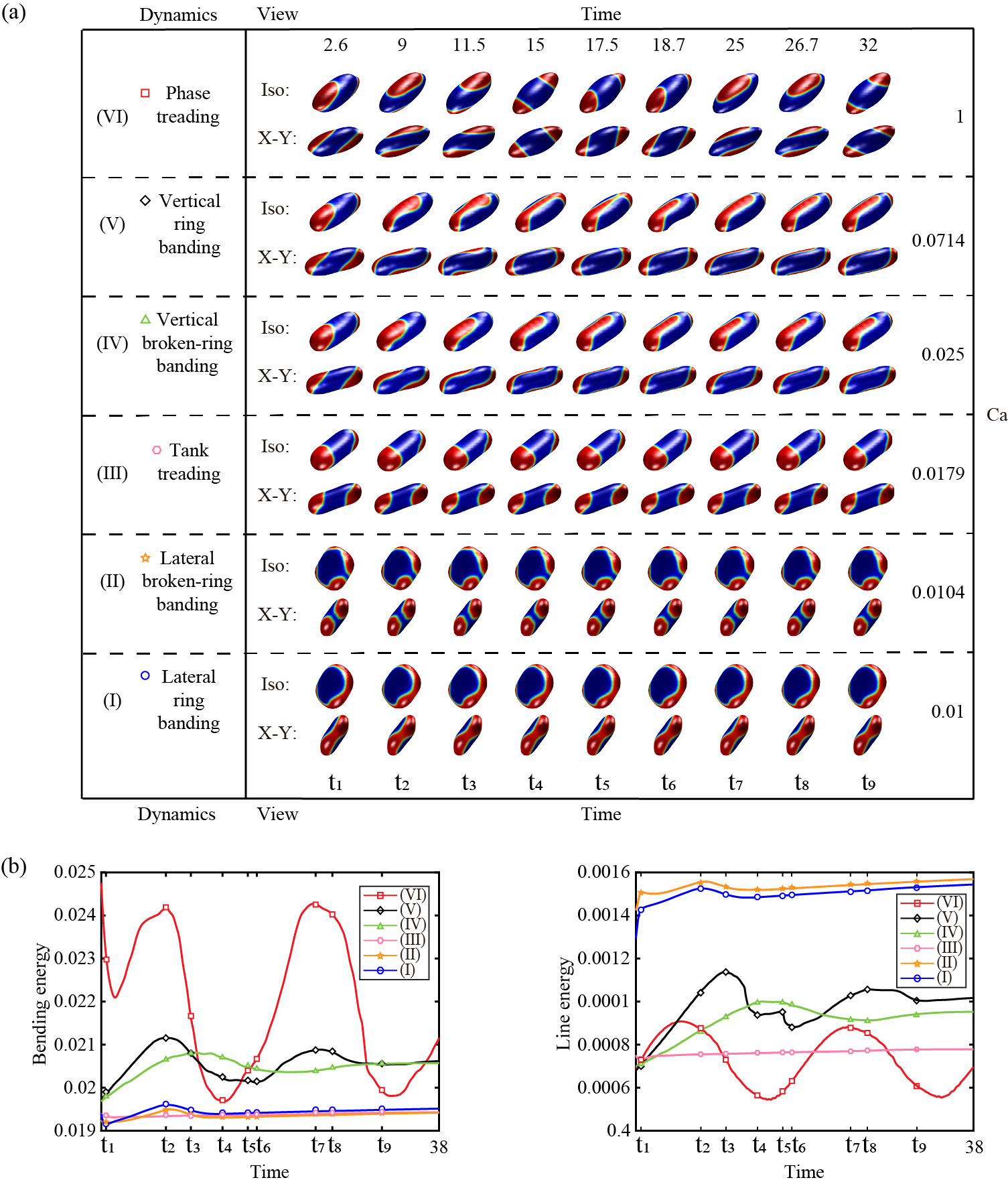}
   \caption{(a) Time evolution of six vesicle dynamic regimes \textbf{(I)}--\textbf{(VI)} at $ \kappa_B = 0.65 $ with varying $Ca$. (b) Corresponding evolution of bending and line energies. Key time points: $t_1 = 2.6$, $t_2 = 9$, $t_3 = 11.5$, $t_4 = 15$, $t_5 = 17.5$, $t_6 = 18.7$, $t_7 = 25$, $t_8 = 26.7$, $t_9 = 32$.}
   \label{fig:kb-ca}
\end{figure}

Figure~\ref{fig:kb-ca} presents a phase diagram spanned by $(\kappa_B, Ca)$, capturing diverse dynamics of multicomponent vesicles and identifying six distinct dynamical regimes: 
\begin{itemize} 
    \item[ \textbf{(I)}] \textbf{Lateral ring banding:} The vesicle adopts an oblate, tilted shape, with \( l_d \) phase domains stretching along the edges to form a lateral ring band composed of a single slender domain. 
    \item[\textbf{(II)}] \textbf{Lateral broken-ring banding:} Similar to \textbf{(I)}, but the \( l_d \) phase domains form a broken lateral ring composed of two slender domains. 
    \item[\textbf{(III)}] \textbf{Tank treading:} The vesicle assumes a prolate, tilted shape, with \( l_d \) phase domains stably located at the high-curvature tips (reported in \cite{liu_dynamics_2017, venkatesh_shape_2024}). 
    \item[\textbf{(IV)}] \textbf{Vertical broken-ring banding:} The \(l_d\) phase domains stretch vertically along the flow direction on the membrane of a prolate, tilted vesicle. This results in a vertical broken ring composed of two slender domains.
    \item[\textbf{(V)}] \textbf{Vertical ring banding:} Similar to \textbf{(IV)}, but the \( l_d \) phase domains merge into a complete vertical ring band composed of a single slender domain (reported in \cite{gera_three-dimensional_2018}).
    \item[\textbf{(VI)}] \textbf{Phase treading:}  The \( l_d \) phase domains migrate periodically along the membrane of a prolate, tilted vesicle from one tip to the other tip in the flow direction, while maintaining the initial domain symmetry (reported in \cite{liu_dynamics_2017, gera_three-dimensional_2018, venkatesh_shape_2024}).
\end{itemize} 
Note that, regimes \textbf{(I)}, \textbf{(II)}, and \textbf{(IV)} have not been previously reported and represent novel dynamical regimes driven by membrane bending heterogeneity under shear flow.

The phase diagram also reveals two prominent ''dynamical belts'':
(i) a ''Low-Rigidity Full-Dynamics Belt'' ($ 0.25 \leq \kappa_B \leq 0.65$), where the membrane rigidity contrast is small enough for all six regimes to appear across a wide range of $Ca$; and (ii) a ''Low-Capillary Full-Dynamics Belt'' ($0.027 \lesssim Ca \lesssim 0.043$), where relatively weak shear allows all six regimes to emerge over a range of $\kappa_B$.

Within the ''Low-Rigidity Full-Dynamics Belt'' ($0.25 \leq \kappa_B \leq 0.65$), increasing $Ca$ drives a flow-mediated transition pathway through the dynamical regimes: at low $Ca$, weak shear and dominant bending energy lead the vesicle to adopt an oblate shape, with low-rigidity $l_d$ domains accumulating along high-curvature lateral edges, forming \textbf{(I) Lateral ring banding}. As $Ca$ increases moderately, enhanced shear stresses partially rupture the $l_d$ domains at the lateral edges, producing \textbf{(II) Lateral broken-ring banding}. With further increase in $Ca$, the vesicle elongates into a prolate shape, and the $l_d$ domains stabilize at the high-curvature tips, resulting in \textbf{(III) Tank treading}. At higher $Ca$, the $l_d$ domains stretch vertically along the membrane, but insufficient membrane fluidity prevents full closure into a ring, resulting in \textbf{(IV) Vertical broken-ring banding}. At sufficiently high $Ca$, the $l_d$ domains are further stretched and coalesce into--\textbf{(V) Vertical ring banding}. Under even stronger $Ca$, persistent and periodic domain migration emerges, characterized by symmetric phase migration, known as \textbf{(VI) Phase treading}.

Outside the ''Low-Rigidity Full-Dynamics Belt'' ($\kappa_B > 0.65$), the diversity of accessible dynamical regimes progressively narrows. For $0.65 \leq \kappa_B <  0.75$, only regimes \textbf{(III)}-\textbf{(VI)} remain, while for $\kappa_B > 0.75$, regime \textbf{(III)} also vanishes. This reduction in dynamical regimes results from the diminishing bending-rigidity contrast between the $l_d$ and $l_o$ phases, which weakens the tendency of low rigidity $l_d$ domains to localize at high-curvature regions and suppresses bending-driven behaviors such as \textbf{(I) lateral ring banding}.

\begin{figure}[h]
   \centering
   \includegraphics[width=0.8\linewidth]{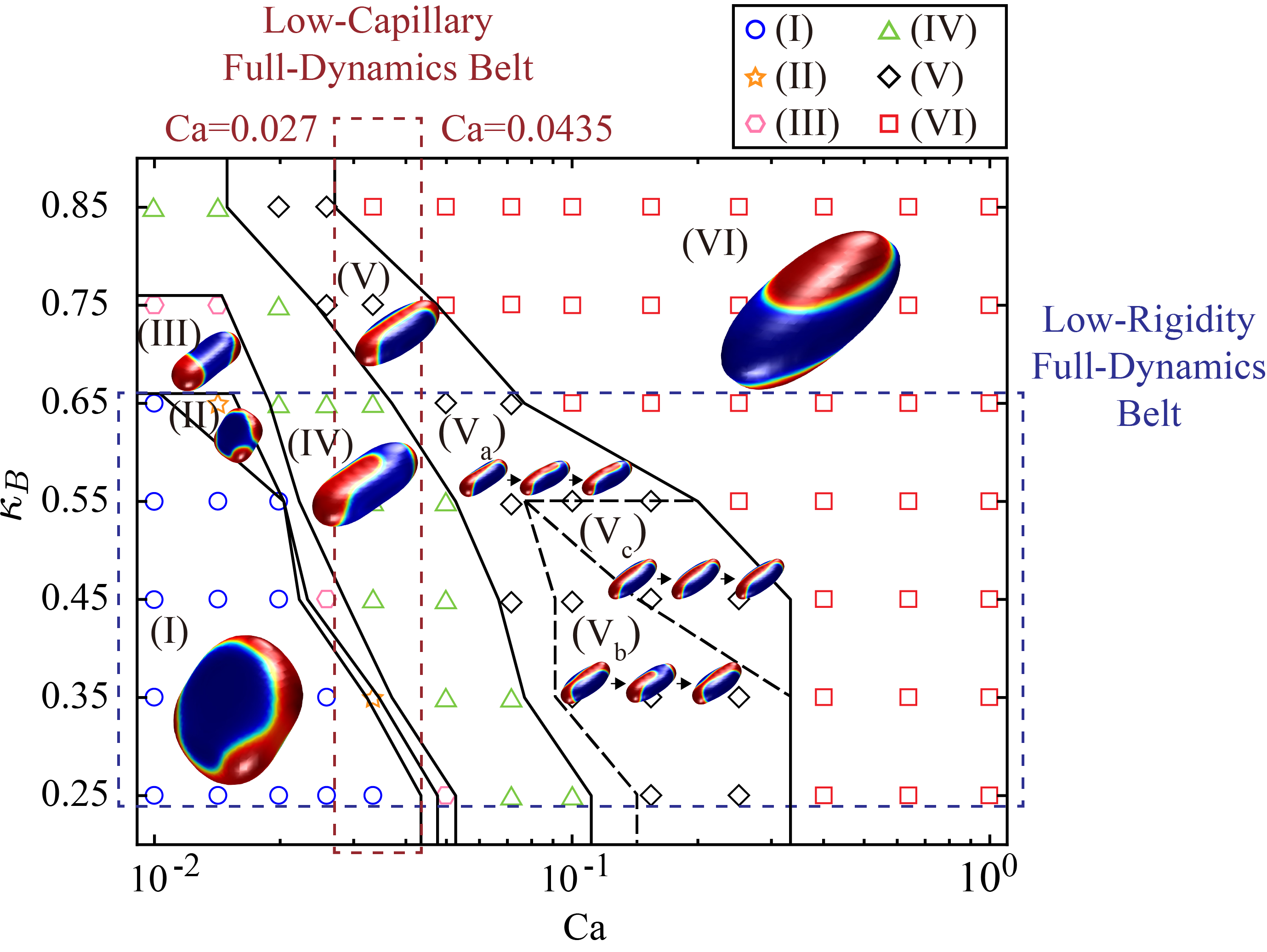}
   \caption{ Phase diagram of vesicle dynamics under shear flow as a function of bending rigidity contrast $\kappa_B$ and capillary number $Ca$. Regimes \textbf{(I)}-\textbf{(VI)} are distinguished by colors and markers.}
   \label{fig:kb-ca-diagram}
\end{figure}

Notably, within the ‘‘Low-Capillary Full-Dynamics Belt’’ ($0.027 \leq Ca \lesssim 0.0435$) shown in figure~\ref{fig:kb-ca-diagram}, we observe a complete sequence of transitions from regime \textbf{(I)} to regime \textbf{(VI)} as $\kappa_B$ increases. This highlights the strong sensitivity of both domain and shape evolution to bending rigidity contrast under moderate shear, and identifies this belt as a critical parameter space for capturing diverse vesicle dynamical regimes.

We further observe that the critical capillary numbers separating successive regimes-for example, the transition from \textbf{(IV)} to \textbf{(V)}-generally decrease with increasing $\kappa_B$, except for the \textbf{(V)} to \textbf{(VI)} transition at $\kappa_B \leq 0.45$. This trend indicates that stronger flow is typically required to induce domain migration and evolution under pronounced bending heterogeneity, reflecting a competition between shear-induced forces and membrane bending rigidity contrast.

Moreover, within the \textbf{(V) Vertical ring banding} regime in figure~\ref{fig:kb-ca-diagram}, we identify three sub-regimes based on ring dynamics: 
\begin{itemize}
     \item[\textbf{(V$_\text{a}$)}] \textbf{Temporary ring:} The $l_d$ domains briefly merge into a vertical ring that ruptures and stabilizes into a broken-ring configuration.
     \item[\textbf{(V$_\text{b}$)}] \textbf{Cyclic broken and closing ring:} The $l_d$ domains repeatedly merge into a vertical ring that undergoes persistent cycles of rupture and reconnection.
     \item[\textbf{(V$_\text{c}$)}] \textbf{Stable ring:} The $l_d$ domains experiences several rupture-reconnect cycles but eventually stabilizes and remains intact over long times.
\end{itemize}
These sub-regimes emerge from transitions driven by the bending heterogeneity $\kappa_B$ and capillary number $Ca$. At $\kappa_B \geq 0.65$, a weak bending heterogeneity reduces the preference of $l_d$ domains for high-curvature tips, restricting dynamics to \textbf{(V$_\text{a}$)}. For $0.45 < \kappa_B < 0.65$, increasing $Ca$ triggers sequential transitions from \textbf{(V$_\text{a}$)} to \textbf{(V$_\text{b}$)} and ultimately to \textbf{(V$\text{c}$)}. At $\kappa_B \leq 0.45$, a strong heterogeneity destabilizes the ring, suppressing \textbf{(V$_\text{c}$)}. Notably, \textbf{(V$_\text{a}$)} is distinct from regime \textbf{(IV)}: in \textbf{(V$_\text{a}$)}, a transient ring forms and then ruptures, whereas in \textbf{(IV)}, the $l_d$ domains never coalesce into a ring. 

Figure~\ref{fig:kb-ca} illustrates six distinct vesicle dynamic regimes \textbf{(I)-(VI)} at $\kappa_B = 0.65$ within the ''Low-Rigidity Full-Dynamics Belt''. Figure~\ref{fig:kb-ca}(a) shows their temporal evolution (see supplementary Movie-1, Movie-2, Movie-3, Movie-4, Movie-5, and Movie-6, corresponding to regimes \textbf{(I)}-\textbf{(VI)}, respectively), while figure~\ref{fig:kb-ca}(b) presents the corresponding bending and line energy profiles from $t = 2$ to $t = 38$. To capture the key dynamical processes, we mark the critical turning points ($t_{i}$ for $i=1,...,9$, see the caption of figure~\ref{fig:kb-ca}) on the energy curves. These turning points of bending and line energies largely coincide.

\begin{figure}[h]
   \centering
   \includegraphics[width=1\linewidth]{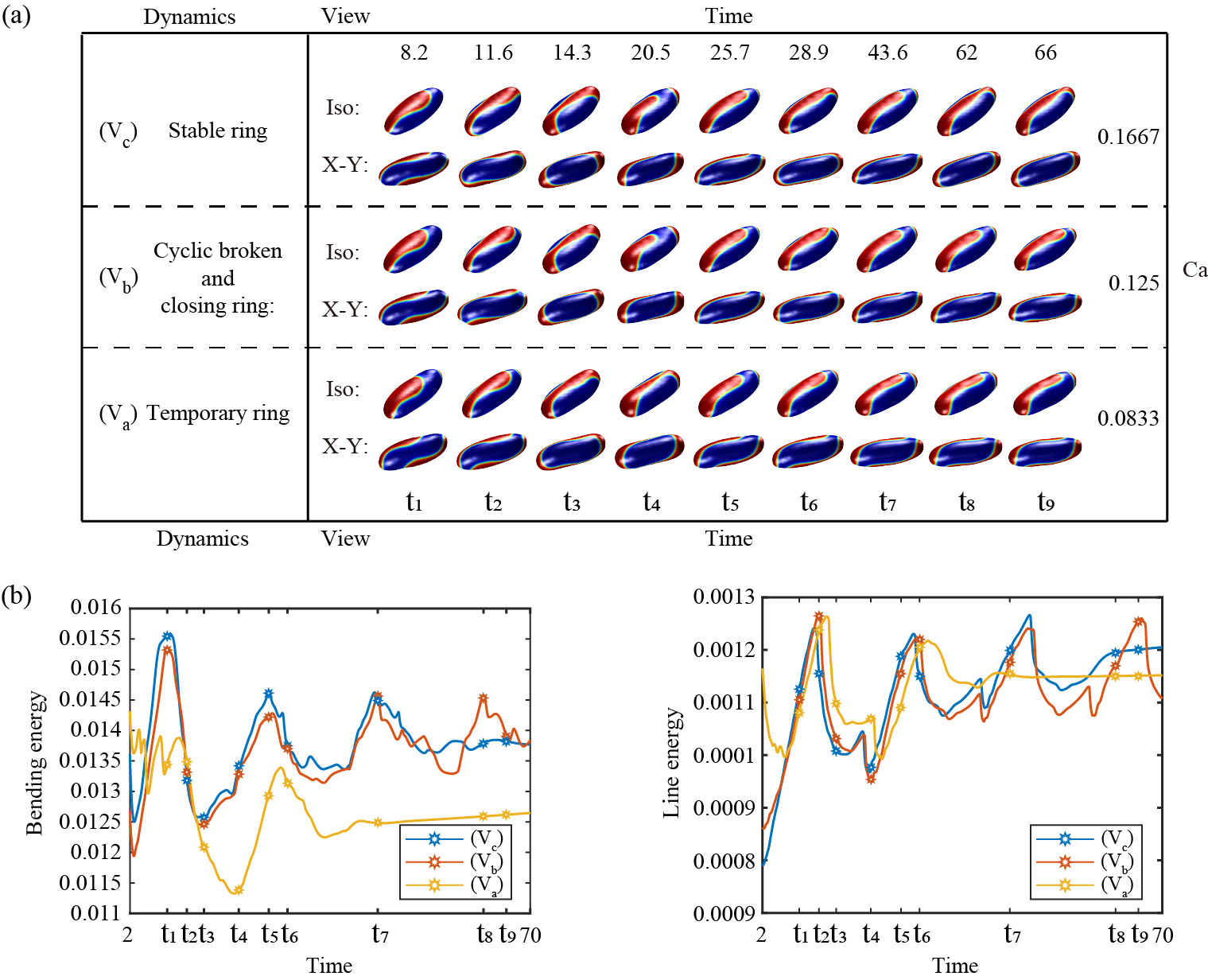}
    \caption{(a) Time evolution of sub-regimes \textbf{(V$_a$)}--\textbf{(V$_c$)} within regime \textbf{(V)} at $ \kappa_B = 0.45 $ with varying $Ca$. (b) Corresponding evolution of bending and line energies. Key time points: $t_1 = 8.2$, $t_2 = 11.6$, $t_3 = 14.3$, $t_4 = 20.5$, $t_5 = 25.7$, $t_6 = 28.9$, $t_7 = 43.6$, $t_8 = 62$, $t_9 = 66$.}
   \label{fig.ring-dynamics}
\end{figure}

Comparing the $ l_d $ phase domain positions in figure~\ref{fig:kb-ca}(a) with the energy evolution in figure~\ref{fig:kb-ca}(b) provides valuable insights into vesicle dynamics.
In regimes \textbf{(I)}--\textbf{(III)}, $l_d$ domains (with lower rigidity $\kappa_{B_{l_d}}$) localize at high-curvature regions, minimizing bending energy and resulting in similar energy levels. However, due to the relatively low $Ca$, bending forces dominate over viscous forces, preventing further reduction in line energy.
In regime \textbf{(IV)}, the $l_d$ domains stretch from the high-curvature tips toward low-curvature regions away from the tips, leading to a temporary increase in bending energy from $t_1$ to $t_3$. Unable to overcome the energy barrier required to form a vertical ring, both the bending and line energies stabilize thereafter at $t_7$. 
In regime \textbf{(V)}, the $l_d$ domains initially stretch toward low-curvature membrane regions away from the tips, increasing bending and line energies from $t_1$ to $t_2$. At $t_4$, the domains merge into a vertical ring band, minimizing both energies. After $t_9$, the domains stabilize into a broken-ring configuration. Strictly, this case at $\kappa_B = 0.65$ corresponds to sub-regime \textbf{(V$_\text{a}$)}, with detailed sub-regime dynamics presented in figure~\ref{fig.ring-dynamics}.
In \textbf{(VI)}, $l_d$ domains periodically migrate along the membrane. Bending energy peaks when domains align along flatter regions (e.g., $t_2$, $t_7$) and decreases when they reach high-curvature tips (e.g., $t_4$, $t_9$).

To further investigate the \textbf{(V) Vertical ring banding} regime, figure~\ref{fig.ring-dynamics} presents three sub-regimes: \textbf{(V$_\text{a}$)}, \textbf{(V$_\text{b}$)}, and \textbf{(V$_\text{c}$)}.
figure~\ref{fig.ring-dynamics}(a) shows the temporal evolution of each dynamics sub-regime, while figure~\ref{fig.ring-dynamics}(b) shows the corresponding bending and line energies from $t=2$ to $t = 70$, with key transition points ($t_i$, $i = 1, \dots, 9$, see the caption of figure~\ref{fig.ring-dynamics}) marked.
Consistent with figure~\ref{fig:kb-ca}, the $l_{d}$ phase domain positions correlate with energy evolution.
We find that the three sub-regimes exhibit similar energy evolution before $t_4$, as all initially form a vertical ring. Subsequent divergence arises from differences in $l_d$ domain evolution.
In \textbf{(V$_\text{a}$)}, the ring ruptures and stabilizes as a broken-ring after $t_7$, with both bending and line energies reaching stability.
In \textbf{(V$_\text{b}$)}, the ring undergoes periodic rupture-reconnection, leading to cyclic oscillations of the energy: bending energy increases during rupture (e.g., $t_3$ to $t_4$) as more low-rigidity $l_d$ domains migrate to low-curvature regions away from the tips, and decreases during reconnection (e.g., $t_5$ to $t_6$).
In \textbf{(V$_\text{c}$)}, after two rupture-reconnection cycles (from $t_3$ to $t_8$), the $l_d$ domains form a stable vertical ring at $t_8$, with both bending and line energies stabilizing thereafter.
Notably, the equilibrium energies in \textbf{(V$_\text{c}$)} are higher than in \textbf{(V$_\text{a}$)} due to (i) a larger fraction of low-rigidity $l_{d}$ domains remaining in low-curvature regions, increasing bending energy, and (ii) a longer $l_d$-$l_o$ interface, elevating line energy.

\subsection{Surface tension heterogeneity driven swinging and tumbling}\label{sec5:ks-ca} \ 

We next investigate the role of membrane surface tension heterogeneity  $\sigma_S = \sigma_{S_{l_d}} / \sigma_{S_{l_o}}$ in modulating vesicle dynamics under shear flow. Within the ''Low-Rigidity Full-Dynamics Belt'' at a fixed bending rigidity contrast \( \kappa_B = 0.65 \), we systematically vary \( \sigma_S \leq 1 \) and the bending capillary number \( Ca \). 

The numerical setup follows Section~\ref{sec.4.1:kb-ca}. In all cases, the $l_d$ phase is assumed to possess lower surface tension than the $l_o$ phase, i.e., $\sigma_{S_{l_d}} < \sigma_{S_{l_o}}$.

\begin{figure}[h]
   \centering
	\includegraphics[width=0.8\textwidth]{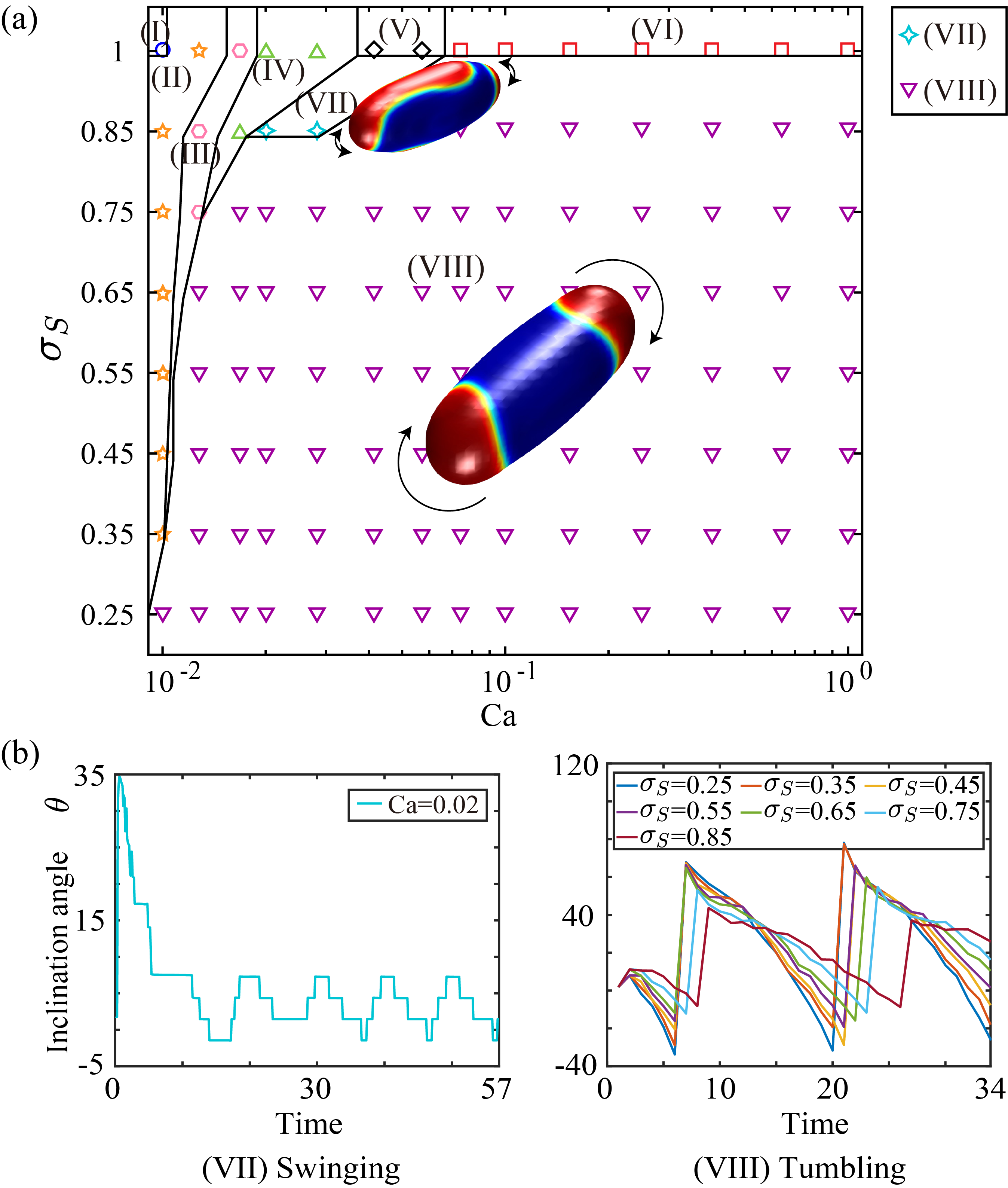}
	\caption{(a) Phase diagram of vesicle dynamics under shear flow as a function of surface tension contrast $\sigma_S$ and capillary number $Ca$ at fixed bending rigidity contrast $\kappa_B = 0.65$. Eight dynamic regimes \textbf{(I)--(VIII)} are identified and distinguished by colors and markers. 
    (b)  Left: inclination angle evolution in regime \textbf{(VII)} at $\sigma_S = 0.85$, $Ca = 0.02$. 
	Right: inclination angle evolution in regime \textbf{(VIII)} at $Ca = 1$ for varying $\sigma_S$.}
	\label{fig:ks-ca}
\end{figure}

The phase diagram (figure~\ref{fig:ks-ca}) reveals eight dynamical regimes: the six previously identified (\textbf{I}-\textbf{VI}) and two additional regimes: 
\begin{itemize}
    \item[\textbf{(VII)}] \textbf{ Swinging:} Small-amplitude oscillations of the vesicle inclination angle, coupled with cyclic rupture and reconnection of the $l_d$-formed vertical ring, similar to \textbf{(V$_\text{b}$)}.
    \item[\textbf{(VIII)}] \textbf{Tumbling:} Rigid-body-like vesicle rotation, with $l_d$ domains stably anchored at high-curvature tips. 
\end{itemize}
Swinging and tumbling dynamics have been previously reported in single-component vesicles with viscosity contrast, and in multicomponent vesicles with small-deformation bending heterogeneity \cite{gera_swinging_2022, venkatesh_shape_2024} or with shape asymmetry arising from initial phase distributions and bending heterogeneity \cite{liu_dynamics_2017}. Here, we identify a novel mechanism: surface tension heterogeneity, without viscosity contrast or asymmetric phase distributions, can also induce swinging and tumbling. As shown in figure~\ref{fig:ks-ca}, regimes \textbf{(VII)} and \textbf{(VIII)} occur only under heterogeneous surface tension ($\sigma_S < 1$). 

Under uniform surface tension (\(\sigma_S = 1\)), moderate capillary numbers (\(0.0189 \lesssim Ca \lesssim 0.037\)) drive the vesicle into regime \textbf{(IV)}, characterized by vertical broken-ring banding. As \(\sigma_S\) decreases to 0.85, enhanced membrane fluidity enables \(l_d\) domains to overcome the energy barrier, forming a complete ring. Periodic rupture and reconnection of this vertical ring dynamically alter the inclination angle, resulting in regime \textbf{(VIII)}. At high capillary numbers (\( 0.037 \lesssim  Ca \leq 1 \)), uniform surface tension (\(\sigma_S = 1\)) induces regimes \textbf{(V)}, characterized by vertical ring banding, and \textbf{(VI)}, defined by phase treading. As \(\sigma_S\) decreases to 0.85, regime \textbf{(VIII) Tumbling} emerges. This transition occurs because the lower-tension \(l_d\) domains become strongly anchored at the high-curvature tips, suppressing their migration and introducing hydrodynamic asymmetry that promotes rigid-body-like rotation.

Overall, decreasing $\sigma_S$ lowers the energetic cost of deforming $l_d$ domains, enhancing their mobility and reducing the critical capillary number $Ca$ required for successive dynamic regime transitions. As a result, at $0.85 \leq \sigma <1$ only regimes \textbf{(II)-(IV)} \textbf{(VII)}, \textbf{(VIII)} persist, regime \textbf{(I)} becomes unsustainable under shear, while swinging and tumbling emerge as the dominant dynamic responses. In addition, for $0.35 \leq \sigma_S <0.85 $, only regimes \textbf{(II)}, \textbf{(III)}, and \textbf{(VIII)} persist. In this range, strong surface tension contrast firmly anchors $l_d$ domains at the tips, promoting vesicle rotation and lowering the threshold for tumbling, while suppressing other regimes as domain evolution rapidly stabilizes. When $\sigma_S < 0.35$, only \textbf{(VIII)} is observed across all tested $Ca$, indicating that extreme surface tension asymmetry robustly favors rigid-body-like rotation.

The distinct signatures of \textbf{(VII) Swinging} and \textbf{(VIII) Tumbling} are further quantified in figure~\ref{fig:ks-ca}(b), which show the time evolution of the vesicle inclination angle for regimes \textbf{(VII)} and \textbf{(VIII)}. In \textbf{(VII) Swinging}, the angle exhibits step-like segments linked to the rupture and reconnection of the $l_d$ ring. As $l_d$ domains stretch toward low-curvature regions, surface tension gradients reduce the inclination angle; after rupture, domains shift back to the tips, increasing the angle. These non-smooth variations reflect the interplay between domain evolution and shear force. In \textbf{(VIII) Tumbling}, further decreasing $\sigma_S$ at a fixed $Ca$ accelerates the angular rotation, driven by stronger surface tension gradients.

\begin{figure}[h]
   \centering
   \includegraphics[width=\linewidth]{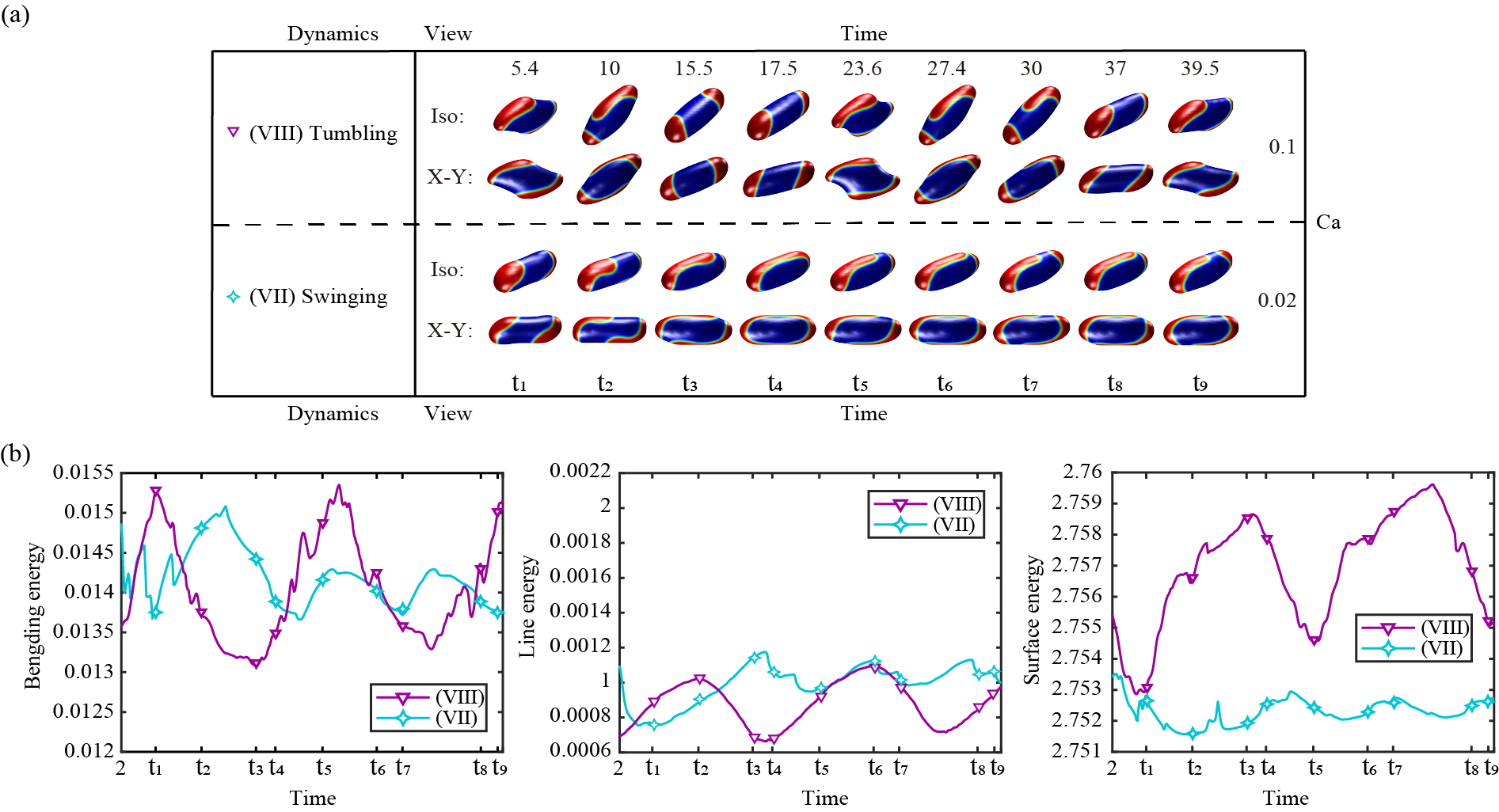}
   \caption{(a) Time evolution of vesicle dynamic regimes \textbf{(VII)} and \textbf{(VIII)} at $ \sigma_S = 0.85 $ and $ \kappa_B = 0.65 $. (b) Corresponding evolution of bending, line, and surface energies. Key time points: $t_1 = 5.4$, $t_2 = 10$, $t_3 = 15.5$, $t_4 = 17.5$, $t_5 = 23.6$, $t_6 = 27.4$, $t_7 = 30$, $t_8 = 37$, $t_9 = 39.5$.}
   \label{fig:ks-ca-dynamics}
\end{figure}

In addition, figure~\ref{fig:ks-ca-dynamics} provides a comprehensive depiction of the \textbf{(VII) Swinging} and \textbf{(VIII) Tumbling} regimes. Figure~\ref{fig:ks-ca-dynamics}(a) illustrates their temporal evolution (see supplementary Movie-7 and Movie-8, corresponding to regimes \textbf{(VII)}-\textbf{(VIII)}, respectively), while figure~\ref{fig:ks-ca-dynamics}(b) presents the corresponding bending, line, and surface energies starting from $t = 2$ to $t = 40$. The critical time points ($t_i$, for $i = 1, \dots, 9$; see caption of figure~\ref{fig:ks-ca-dynamics}) are marked on the energy curves. The evolution of bending and line energies, which reflects the migration of $l_d$ domains, is consistent with the discussion in Section~\ref{sec.4.1:kb-ca} and is not repeated here. The following analysis focuses on the variations in surface energy.

In \textbf{(VII) Swinging}, as shown in figure~\ref{fig:ks-ca-dynamics}(a), periodic rupture and reconnection of the ring cause surface energy to oscillate. When the ring reconnects, $l_d$ domains stretch into low-curvature regions, increasing both their area and the surface energy (e.g., around $t_4$). When the ring ruptures, $l_d$ domains cluster near the high-curvature tips, reducing their area and lowering surface energy (e.g., around $t_5$).

In \textbf{(VIII) Tumbling}, the $l_d$ domains remain anchored at the tips throughout the vesicle’s rotation. The vesicle alternates between elongated shapes, tilted along the flow direction (with inclination angle $\theta \in (0, \pi/2)$), and flattened shapes, tilted against the flow ($\theta \in (-\pi/2, 0)$). When elongated (e.g., at $t_3$), the vesicle adopts a thin, prolate shape, where the $l_o$ domains are located in relatively low-curvature regions. When flattened (e.g., at $t_5$), the vesicle deforms into a more oblate shape under shear flow, causing the curvature at the $l_o$ locations to decrease further. This curvature reduction lowers the surface energy. Such cyclic shape evolution, driven by the interplay between shear forces and surface tension heterogeneity, accounts for the periodic surface energy variations observed in the tumbling regime.

\section{Conclusion} \

We have developed a comprehensive, thermodynamically consistent phase-field model that accurately captures the intricate interplay among vesicle dynamics, membrane heterogeneity, and shear flow. The model successfully reproduces experimental observations of lateral phase separation and shear-induced morphological transitions. This underscores the accuracy, versatility, and predictive capability of the model and numerical methods.

Systematic variation of the bending rigidity contrast $\kappa_B$ and capillary number $Ca$ reveals six dynamic regimes, including three previously unreported: \textbf{(I)}, \textbf{(II)}, and \textbf{(IV)}. These regimes organize into two critical regions--the ''Low-Rigidity Full-Dynamics Belt'' and the ''Low-Capillary Full-Dynamics Belt''--where all six dynamical regimes coexist. Within regime \textbf{(V)}, we further identify three sub-regimes that reflect the rich ring structure of domain redistribution under shear. Beyond bending effects, we uncover surface tension heterogeneity as a previously unknown mechanism that suffices to induce \textbf{(VII)} and \textbf{(VIII)}, revealing a new pathway for vesicle dynamics under shear. This unveils a distinct physical mechanism for tuning vesicle dynamics.

These findings offer a comprehensive understanding of the role of membrane heterogeneity in governing vesicle dynamics under shear flow, providing new insights and design principles for tunable vesicle-based carriers.

%\appendix
 \newpage
	\appendix	
	\section{Diffuse domain method}\label{appendix-ddm}
    \renewcommand{\thesection}{\Alph{section}} % 附录章节编号为 A, B, ...
    \renewcommand{\theequation}{\thesection.\arabic{equation}} % 公式编号为 A.x
   \setcounter{equation}{0} % 重置公式编号
   \

We describe the diffuse domain method used to extend membrane dynamics from the sharp interface $\Gamma$ to the rectangular domain $\Omega$. This extension is achieved through a surface delta function $\delta(\phi)$ \eqref{delta}, constructed from the phase-field variable $\phi$, which is initialized with a hyperbolic tangent profile \eqref{eq:phi initial} to characterize the diffuse vesicle.
The surface delta function $\delta(\phi)$ satisfies the following approximation membrane~\cite{teigen2011diffuse}:
\begin{equation} \label{eq:delta relation}
    \int_{\Gamma} w \, \mathrm{d}S \approx \int_{\Omega} w \zeta(\phi) \, \mathrm{d}V,
\end{equation}
for any scalar or vector field $w$ (e.g., $c_{\Gamma}$ or $\boldsymbol{q}_c$). 
This relation enables the approximation of surface integrals and surface differential operators by volume integrals over $\Omega$, facilitating the extension of surface quantities, equations, and energy functionals into $\Omega$.

We now describe the detailed procedure for coupling the surface phase separation dynamics, governed by the surface CH equation on $\Gamma$ \eqref{ch_vs_sharp} to $\Omega$. Using the diffuse domain method \cite{teigen2009diffuse, li2009solving, ratz_pdes_2006}, we assume that the membrane component  $c_{\Gamma}$ is smoothly extended off $ \Gamma $, remaining constant in the normal direction within a narrow band around the membrane. This yields the normal extension condition:
\begin{align} \label{normal_grad}
    \boldsymbol{n} (\boldsymbol{n} \cdot \boldsymbol{\nabla} c_{\Gamma}) = \boldsymbol{0}.
\end{align} 
Additionally, since the diffusive flux $\boldsymbol{q}_c$ is tangential to $\Gamma$, its extension satisfies:
\begin{align} 
    \label{normal_qc}
    \boldsymbol{n} \cdot \boldsymbol{q_c} &= 0.
\end{align}
Using the incompressibility of the background fluid \eqref{incompress_un} and the surface gradient operator defined in \eqref{eq:surface gradient operator}, we derive the following identity near the vesicle membrane $\Gamma$:
\begin{align} \label{div_with_u}
    \boldsymbol{{\nabla}_{\Gamma}} \cdot (c_{\Gamma} \boldsymbol{u}) &= \left( \boldsymbol{\nabla} \cdot \boldsymbol{u} - \boldsymbol{n} \cdot \boldsymbol{\nabla} \boldsymbol{u} \cdot \boldsymbol{n} \right) c_{\Gamma} + \boldsymbol{u} \cdot \left( \boldsymbol{\nabla} c_{\Gamma} - \boldsymbol{n} (\boldsymbol{n} \cdot \boldsymbol{\nabla} c_{\Gamma}) \right) \\ \nonumber
    &= - \left( \boldsymbol{n} \cdot \boldsymbol{\nabla} \boldsymbol{u} \cdot \boldsymbol{n} \right) c_{\Gamma} + \boldsymbol{u} \cdot \left( \boldsymbol{\nabla} c_{\Gamma} - \boldsymbol{n} (\boldsymbol{n} \cdot \boldsymbol{\nabla} c_{\Gamma}) \right).
\end{align}
% where $ \boldsymbol{{\nabla}_{\Gamma}} \cdot \boldsymbol{n} $ denotes the total curvature.
Applying the identity \eqref{div_with_u} to the surface CH equation \eqref{ch_vs_sharp} yields:
\begin{align}
    &(c_{\Gamma})_t - (\boldsymbol{n} \cdot (\boldsymbol{\nabla} \boldsymbol{u}) \cdot \boldsymbol{n}) c_{\Gamma} + \boldsymbol{u} \cdot (\boldsymbol{\nabla} c_{\Gamma} - \boldsymbol{n} (\boldsymbol{n} \cdot \boldsymbol{\nabla} c_{\Gamma}))  = \boldsymbol{\nabla} \cdot \boldsymbol{q_c} - (\boldsymbol{n} \cdot \boldsymbol{\nabla}) (\boldsymbol{n} \cdot \boldsymbol{q_c}). \label{ch_vs_un_extend_A}
\end{align}
Moreover, we further employ the identity from~\cite{james2004surfactant}:
\begin{align} \label{deltaphi eq}
\partial_t \zeta(\phi) + \boldsymbol{u} \cdot \boldsymbol{\nabla} \zeta(\phi) = -(\boldsymbol{n} \cdot \boldsymbol{\nabla} \boldsymbol{u} \cdot \boldsymbol{n}) \zeta(\phi).
\end{align}
Multiplying \eqref{ch_vs_un_extend_A} by $ \zeta(\phi) $, and \eqref{deltaphi eq} by $ c_{\Gamma} $, with the help of the assumptions \eqref{normal_grad} and \eqref{normal_qc}, we obtain
\begin{align} \label{ch_vs_un_ddm_A}
    {(\zeta(\phi) c_{\Gamma})}_t + \boldsymbol{\nabla} \cdot (\zeta(\phi) c_{\Gamma} \boldsymbol{u}) &= \zeta(\phi) \boldsymbol{\nabla} \cdot \boldsymbol{q_c}.
\end{align}
For the flux term, noting that $\zeta(\phi)$ varies predominantly in the normal direction near $\Gamma$ with negligible tangential variation (especially for thin interfaces) and applying \eqref{normal_qc}, we obtain
\begin{align} \label{incompress_vs_un_ddm_A}
    \boldsymbol{\nabla} \cdot (\zeta(\phi) \boldsymbol{q_c}) &= \zeta(\phi) \boldsymbol{\nabla} \cdot \boldsymbol{q_c} + \boldsymbol{\nabla} \zeta(\phi) \cdot \boldsymbol{q_c} \\ \nonumber
    &= \zeta(\phi) \boldsymbol{\nabla} \cdot \boldsymbol{q_c} + \boldsymbol{{\nabla}_{\Gamma}} \zeta(\phi) \cdot \boldsymbol{q_c} + (\boldsymbol{n} \cdot \boldsymbol{\nabla} \zeta(\phi)) (\boldsymbol{n} \cdot \boldsymbol{q_c}) \\ \nonumber
    &= \zeta(\phi) \boldsymbol{\nabla} \cdot \boldsymbol{q_c}.
\end{align}

Thus, the resulting surface CH equation in diffuse domain formulation within $ \Omega $ can be written as:
	\begin{align} 
		\label{ch_vs_un_ddm_AAA}
		 \left(\zeta(\phi) c_{\Gamma}\right)_t + \boldsymbol{\nabla} \cdot  \left(\zeta(\phi) c_{\Gamma} \boldsymbol{u}   \right) &=  \boldsymbol{\nabla} \cdot\left( \zeta(\phi) \boldsymbol{q_c} \right).
	\end{align}

Notably, previous studies have demonstrated that in the diffuse domain method, the extended PDE \eqref{ch_vs_un_ddm_AAA} asymptotically converges to the original surface PDE \eqref{ch_vs_sharp} as the diffuse interface thickness approaches zero, $\varepsilon_{\phi}  \to 0$ \cite{li2009solving, yu2012extended, guo_diffuse_2021, guo_nonlinear_2023, lervag_analysis_2015, chadwick2018numerical, poulsen2018smoothed, yu2020higher}.

\section{Model derivation}\ \label{appendix-modle} \

Next, we derive the free energy dissipation law and the governing equations for the coupled multicomponent vesicle--fluid system. In accordance with the Second Law of Thermodynamics, the free energy $\mathcal{F}$ \eqref{eq:energy} must satisfy the following dissipation law:
	\begin{align} \label{dissipationE}
		\frac{d \mathcal{F}}{d t} & = \int_{\Omega} \frac{\delta\mathcal{F}}{\delta \boldsymbol{u}} \boldsymbol{u}_t +  \frac{\delta\mathcal{F}}{\delta \phi} {\phi}_t+  \frac{\delta\mathcal{F}}{\delta c_{\Gamma}} (c_{\Gamma})_t \, \mathrm{d} V \leq 0, \\ \nonumber
	\end{align}
 where ${\delta\mathcal{F}}/{\delta \boldsymbol{u}}$, ${\delta\mathcal{F}}/{\delta \phi}$, and ${\delta\mathcal{F}}/{\delta c_{\Gamma}}$ denote the variational derivatives of the free energy $\mathcal{F}$ with respect to $\boldsymbol{u}$, $\phi$, and $c_{\Gamma}$, respectively.

For the fluid velocity field $\boldsymbol{u}$, the variational derivative of the free energy $\mathcal{F}$ \eqref{eq:energy} with respect to $\boldsymbol{u}$ is combined with the momentum conservation equation \eqref{ns_un}, applying integration by parts, we obtain
\begin{align}\label{eu}
  	\int_{\Omega} \frac{\delta\mathcal{F}}{\delta \boldsymbol{u}} \boldsymbol{u}_t \, \mathrm{d} V 
  	&= \int_{\Omega} \frac{\delta\mathcal{F}_\text{K}}{\delta \boldsymbol{u}}\boldsymbol{u}_t \, \mathrm{d} V =  \int_{\Omega}\rho \boldsymbol{u}  \cdot \boldsymbol{u}_t\, \mathrm{d} V= \int_{\Omega} -\mu | \boldsymbol{D}(\boldsymbol{u})|^2 + \boldsymbol{F}  \cdot \boldsymbol{u} -\xi \varepsilon_\phi^{2} \phi^{2}\left|\boldsymbol{\nabla} \lambda_{local}\right|^{2}   \, \mathrm{d} V. 
  \end{align}

Similarly, for the phase-field variable $\phi$, the variational derivative of $\mathcal{F}$ with respect to $\phi$ yields:
\begin{align}\label{eq:delta/phi}
 	\frac{\delta\mathcal{F}}{\delta \phi} & = \frac{\delta\mathcal{F}_\text{L}}{\delta \phi} + \frac{\delta\mathcal{F}_\text{S}}{\delta \phi}+\frac{\delta\mathcal{F}_\text{B}}{\delta \phi}   + \frac{\delta\mathcal{F}_\text{A}}{\delta \phi} \\ \nonumber
 	&=\left(   \sigma_L \delta(c_{\Gamma})  + {\sigma_S(c_{\Gamma})}  + M_{s} \frac{S(\phi)-S(\phi_0)}{S(\phi_0)}    \right) \omega_{\phi}  \\ \nonumber
 	&\quad  - {\eta_\phi} {\varepsilon_\phi}\boldsymbol{\nabla}\left(   \sigma_L \delta(c_{\Gamma}) + {\sigma_S(c_{\Gamma}) + M_{s} \frac{S(\phi)-S(\phi_0)}{S(\phi_0)}}   \right) \cdot \boldsymbol{\nabla} \phi  +{\kappa_B(c_{\Gamma})}{\eta_\phi} \frac{f^{\prime \prime}(\phi)}{{\varepsilon_\phi}^2}   \omega_{\phi}   - {\eta_\phi}  \Delta \left( {\kappa_B(c_{\Gamma})} \omega_{\phi}\right) , 
 \end{align}
  where
     \begin{align}
    \omega_\phi &= \eta_\phi \left( \frac{f'(\phi)}{\varepsilon_\phi} - \varepsilon_\phi \Delta \phi \right).
\end{align}

Combining \eqref{eq:delta/phi} with the CH evolution equation for the vesicle membrane \eqref{ch_un} and applying integration by parts, we obtain:
 		\begin{align} \label{ephi}
    	&\int_{\Omega}  \frac{\delta\mathcal{F}}{\delta \phi} {\phi}_t \, \mathrm{d} V  \\ &=\int_{\Omega}  - \boldsymbol{q_{\phi}} \cdot \boldsymbol{\nabla}   \left[\left(   b +  \omega_{c} c_{\Gamma}  \right) \omega_{\phi}  - {\eta_\phi} {\varepsilon_\phi} \boldsymbol{\nabla}\left(  b +  \omega_{c} c_{\Gamma} \right) \cdot \boldsymbol{\nabla} \phi  +{\kappa_B(c_{\Gamma})}{\eta_\phi} \frac{f^{\prime \prime}(\phi)}{{\varepsilon_\phi}^2}   \omega_{\phi}  - {\eta_\phi}  \Delta \left( {\kappa_B(c_{\Gamma})} \omega_{\phi}\right) \right] \nonumber \\ \nonumber
    	&\quad -\boldsymbol{u} \cdot  \boldsymbol{\nabla} \phi \left[\left(   b + \omega_{c} c_{\Gamma}   \right) \omega_{\phi} - {\eta_\phi} {\varepsilon_\phi} \boldsymbol{\nabla}\left(  b +  \omega_{c} c_{\Gamma} \right) \cdot \boldsymbol{\nabla} \phi   +{\kappa_B(c_{\Gamma})}{\eta_\phi} \frac{f^{\prime \prime}(\phi)}{{\varepsilon_\phi}^2}   \omega_{\phi}  - {\eta_\phi}  \Delta \left( {\kappa_B(c_{\Gamma})} \omega_{\phi}\right)  \right]  \, \mathrm{d} V ,   
 \end{align}
where
 \begin{align} 
	b &=  \sigma_L \delta(c_{\Gamma}) + {\sigma_S(c_{\Gamma})}  + M_{s} \frac{S(\phi)-S(\phi_0)}{S(\phi_0)}  -\omega_{c} c_{\Gamma}.
    \end{align}

For the membrane component field $c_\Gamma$, multiplying both sides of \eqref{ch_vs_un} by the chemical potential $\omega_c$, defined as the variational derivative of $\mathcal{F}$ with respect to $c_\Gamma$:
\begin{align}\label{varcgamma}
	\zeta(\phi)\omega_c= \frac{\delta\mathcal{F}}{\delta c_{\Gamma}} & = \frac{\delta\mathcal{F}_\text{L}}{\delta c_{\Gamma}}+\frac{\delta\mathcal{F}_\text{S}}{\delta c_{\Gamma}} + \frac{\delta\mathcal{F}_\text{B}}{\delta c_{\Gamma}} \\ \nonumber
	&=  \frac{{\varepsilon_\phi}^{-1}}{2}{\kappa_B}^{\prime}(c_{\Gamma}) {\omega_{\phi}}^{2}  +  {\sigma_S}^{\prime}(c_{\Gamma}) {\zeta(\phi)}  + \sigma_L \zeta(\phi)\left({\eta_c}  \frac{f^{\prime}(c_{\Gamma})}{\varepsilon_c}-{\eta_c} {\varepsilon_c} \Delta c_{\Gamma}\right)- \sigma_L {\eta_c} {\varepsilon_c} \boldsymbol{\nabla} \zeta(\phi) \cdot \boldsymbol{\nabla}{c_{\Gamma}},
\end{align}
we obtain
\begin{align}
	{\omega_{c}} {(\zeta(\phi) c_{\Gamma})}_t +  {\omega_{c}} \boldsymbol{\nabla} \cdot( \zeta(\phi) c_{\Gamma} \boldsymbol{u})&=  {\omega_{c}} \boldsymbol{\nabla} \cdot\left( \zeta(\phi)  \boldsymbol{q_c} \right).
\end{align}
Using the definition of the chemical potential $\omega_c$ \eqref{varcgamma}, \eqref{ch_un} and \eqref{ch_vs_un}, which describe membrane evolution and the component dynamics of phase separation, respectively, we derive:
  	\begin{align} \label{ecgamma}
  	\int_{\Omega}  \frac{\delta\mathcal{F}}{\delta c_{\Gamma}}  (c_{\Gamma})_t \, \mathrm{d} V & = \int_{\Omega}  {\omega_{c}} \boldsymbol{\nabla} \cdot \boldsymbol{j}-{\omega_{c}} c_{\Gamma} ({ \zeta(\phi)})_t \, \mathrm{d} V \\ \nonumber
  	& = \int_{\Omega} {\omega_{c}} \boldsymbol{\nabla} \cdot \boldsymbol{j}- {{\phi}_t}  \left({\omega_{c}} c_{\Gamma} {\omega_{\phi}}-\eta_{\phi} {\varepsilon_{\phi}}  \boldsymbol{\nabla}\left({\omega_{c}} c_{\Gamma}\right) \cdot \boldsymbol{\nabla} \phi\right) \, \mathrm{d} V \\ \nonumber
  	& = \int_{\Omega} {{\phi}_t}  \left(-{\omega_{c}} c_{\Gamma} {\omega_{\phi}}+\eta_{\phi} {\varepsilon_{\phi}}   \boldsymbol{\nabla}\left({\omega_{c}} c_{\Gamma}\right) \cdot \boldsymbol{\nabla} \phi\right) -\zeta(\phi) \boldsymbol{\nabla} \omega_{c} \cdot \boldsymbol{q_c} +\zeta(\phi) c_{\Gamma} \boldsymbol{\nabla} \omega_{c} \cdot \boldsymbol{u}  \, \mathrm{d} V  \\ \nonumber
  	& = \int_{\Omega}  - \boldsymbol{q_{\phi}} \cdot \boldsymbol{\nabla}  \left(-{\omega_{c}} c_{\Gamma} {\omega_{\phi}}+\eta_{\phi} {\varepsilon_{\phi}}  \boldsymbol{\nabla}\left({\omega_{c}} c_{\Gamma}\right) \cdot \boldsymbol{\nabla} \phi\right) -\zeta(\phi) \boldsymbol{\nabla} \omega_{c} \cdot \boldsymbol{q_c}   \\ \nonumber
  	&\quad \quad -\boldsymbol{u} \cdot \left[ \boldsymbol{\nabla} \phi \left(-{\omega_{c}} c_{\Gamma} {\omega_{\phi}}+\eta_{\phi} {\varepsilon_{\phi}}  \boldsymbol{\nabla}\left({\omega_{c}} c_{\Gamma}\right) \cdot \boldsymbol{\nabla} \phi\right)  - \zeta(\phi) c_{\Gamma} \boldsymbol{\nabla} \omega_c   \right] \, \mathrm{d} V, 
  \end{align}
  where $\boldsymbol{j}=  \zeta(\phi)\boldsymbol{{q_c}} -  \zeta(\phi) c_{\Gamma} \boldsymbol{u} $.

Finally, by substituting \eqref{eu}, \eqref{ephi}, and \eqref{ecgamma} into the free energy dissipation law \eqref{dissipationE}, we obtain
  \begin{align}  \label{eq:energy diss middle}
\frac{d \mathcal{F}}{d t}= & \int_{\Omega}  \frac{\delta \mathcal{F}}{\delta \boldsymbol{u}} { \boldsymbol{u}}_{t}  + \frac{\delta \mathcal{F}}{\delta \phi} {\phi}_{t}+  \frac{\delta \mathcal{F}}{\delta c_{\Gamma}}  (c_{\Gamma})_{t} \, \mathrm{d} V \\ \nonumber
= & \int_{\Omega}-\boldsymbol{q}_{\boldsymbol{\phi}} \cdot \boldsymbol{\nabla}\left[ b\omega_{\phi}       
    + {\kappa_B(c_{\Gamma})}{\eta_\phi} \frac{f^{\prime \prime}(\phi)}{{\varepsilon_\phi}^2}   \omega_{\phi}    -  {\eta_\phi} \Delta({\kappa_B(c_{\Gamma})}  \omega_{\phi} )     - {\eta_\phi} {\varepsilon_\phi} \boldsymbol{\nabla} b   \cdot \boldsymbol{\nabla} \phi  \right] \\ \nonumber
    & -\boldsymbol{u} \cdot\left[\boldsymbol { \nabla } \phi \left[ b\omega_{\phi}  + {\kappa_B(c_{\Gamma})}{\eta_\phi} \frac{f^{\prime \prime}(\phi)}{{\varepsilon_\phi}^2}   \omega_{\phi}    - {\eta_\phi} \Delta({\kappa_B(c_{\Gamma})}  \omega_{\phi} )    - {\eta_\phi} {\varepsilon_\phi} \boldsymbol{\nabla} b   \cdot \boldsymbol{\nabla} \phi    -\zeta(\phi) c_{\Gamma} \boldsymbol{\nabla} \omega_{c}\right]\right] \\ \nonumber
    & - \mu\left|\nabla \boldsymbol{u}\right|^{2} + \boldsymbol{F} \cdot \boldsymbol{u}-\xi \varepsilon_{\phi}^{2} \phi^{2}\left|\nabla \lambda_{local}\right|^{2}  -\zeta(\phi) \boldsymbol{\nabla} \omega_{c} \cdot \boldsymbol{q}_{\boldsymbol{c}} \, \mathrm{d} V \leq 0. \nonumber
\end{align}  

From these relations, we derive explicit expressions for the previously undetermined terms $\boldsymbol{F}$, $\boldsymbol{q}_\phi$, and $\boldsymbol{q}_c$ in \eqref{ns_un}--\eqref{inextend_un} and \eqref{ch_vs_un}:
\begin{align}  \label{eq:F} 
	\boldsymbol{F}&=  g \boldsymbol{\nabla} \phi  -  \zeta(\phi) c_{\Gamma} \boldsymbol{\nabla} \omega_c, \\
	\boldsymbol{q_{\phi}} &=  m(\phi) \boldsymbol{\nabla} {g}, \\ \label{eq:qc}
	\boldsymbol{q_c} &=  m(c_{\Gamma}) \boldsymbol{\nabla} {\omega_c},  
\end{align}
where
\begin{align} \label{eq:g}
	g&=     b\omega_{\phi}       
    + \left({\kappa_B(c_{\Gamma})}{\eta_\phi} \frac{f^{\prime \prime}(\phi)}{{\varepsilon_\phi}^2}   \omega_{\phi}    -  {\eta_\phi} \Delta({\kappa_B(c_{\Gamma})}  \omega_{\phi} )   \right)    - {\eta_\phi} {\varepsilon_\phi} \boldsymbol{\nabla} b   \cdot \boldsymbol{\nabla} \phi, \\ \label{eq:b}
	b &= -\omega_{c} c_{\Gamma}  + \sigma_L \delta(c_{\Gamma})  +  \mathrm{M}_{\text{s}} \frac{S(\phi)-S(\phi_0)}{S(\phi_0)}  + Cs_{\Gamma} {\sigma_S(c_{\Gamma})}.
    \end{align}
Here, the functions $m(\phi)=D_\phi\left(1 - \phi ^2\right)^2$ and $m(c_\Gamma) = D_c\left(1 - c_\Gamma ^2\right)^2$ denote the mobilities, which are phenomenological parameters with $D_\phi$ and $D_c$ being the diffusion coefficients.

By substituting \eqref{eq:F}--\eqref{eq:b} into \eqref{eq:energy diss middle}, we obtain the following free energy dissipation identity:
\begin{align}
    \frac{d \mathcal{F}}{d t}= & \int_{\Omega}  \frac{\delta \mathcal{F}}{\delta \boldsymbol{u}} { \boldsymbol{u}}_{t} + \frac{\delta \mathcal{F}}{\delta \phi} {\phi}_{t}+  \frac{\delta \mathcal{F}}{\delta c_{\Gamma}}  (c_{\Gamma})_{t} \, \mathrm{d} V \\ \nonumber
= & \int_{\Omega} -\mu\left|\nabla \boldsymbol{u}\right|^{2}  -m(\phi)|\boldsymbol{q}_{\boldsymbol{\phi}}|^2  -\zeta(\phi) m(c_\Gamma)|\boldsymbol{q}_{\boldsymbol{c}}|^2 -\xi \varepsilon_{\phi}^{2} \phi^{2}\left|\nabla \lambda_{local}\right|^{2} \leq 0.
\end{align}
This expression confirms the thermodynamic consistency of the model, with free energy monotonically decreasing due to viscous dissipation, diffusive fluxes, and membrane tension relaxation.

\section{ Numerical method} \label{appendix-numerical method} \

To solve the coupled system numerically, we discretize the time interval $I = [0, T]$ into uniform time steps $\Delta t = t_{n+1} - t_n$. Time-dependent variables at the $n$-th and $(n+1)$-th time steps are denoted as $(\cdot)^n$ and $(\cdot)^{n+1}$, respectively.

At each time step, the solution procedure consists of the following sequential steps:

1. We first solve the CH equation \eqref{ch} governing the membrane evolution to update $\phi^{n+1}$, $g^{n+1}$ and $\omega_{\phi}^{n+1}$ by the following method:
\begin{align}\label{phi-method}
&\frac{\phi^{n+1} - \phi^{n}}{\Delta t} +  \boldsymbol{\nabla} \cdot (\phi^{n} \boldsymbol{u}^{n})  = \frac{1}{Pe}\boldsymbol{\nabla} \cdot \boldsymbol{q_{\phi}}^{n+1},  \\ 
&\boldsymbol{q_{\phi}}^{n+1} = m(\phi^{n}) \boldsymbol{\nabla} g^{n+1},  \\
\label{g-method}
&g^{n+1}=   b^{n} \omega_{\phi}^{n+1}  +  \frac{1}{Ca} \left( {\kappa_B(c_{\Gamma}^{n})} {\eta_\phi} \frac{f^{\prime \prime}(\phi^{n})}{{\varepsilon_\phi}^{2}} \omega_{\phi}^{n+1} -  {\eta_\phi} \Delta \left({\kappa_B(c_{\Gamma}^{n})} \omega_{\phi}^{n+1} \right)  \right)  - {\eta_{\phi}} {\varepsilon_{\phi}} \boldsymbol{\nabla} b^{n}   \cdot \boldsymbol{\nabla} \phi^{n}   ,\\
& b^{n} = -\omega_{c}^{n} c_{\Gamma}^{n}  + {Cn_{\Gamma}} \sigma_L\delta(c_{\Gamma}^{n})  +  \mathcal{M}_{\text{s}} \frac{S(\phi^{n})-S(\phi_0)}{S(\phi_0)}  +  {Cs_{\Gamma}}{\sigma_S(c_{\Gamma}^{n})} ,\\ \label{omegaphi-method}
&\omega_{\phi}^{n+1} = {\eta_{\phi}} \left( \frac{f^{\prime}({\phi}^{n+1})}{\varepsilon_{\phi}}- {\varepsilon_{\phi}} \Delta \phi ^{n+1} \right).   
\end{align}
The nonlinear terms, such as $f^{\prime}(\phi^{n+1})$ and $f^{\prime\prime}(\phi^{n+1})$, are linearized via a first-order Taylor expansion, as described in \cite{lowengrub_numerical_2016}.

% \begin{align} \label{taylor}
% f^{\prime}(\phi^{n+1}) = f^{\prime}(\phi^{n}) + f^{\prime\prime}(\phi^{n}) \left( \phi^{n+1} - \phi^{n} \right)
% \end{align}

2. Next, we solve the CH equation \eqref{ch_vs} governing the diffusion of membrane components to update $c_{\Gamma}^{n+1}$ and $\omega_c^{n+1}$ by the following method:
\begin{align} \label{c-method}
&\frac{ \zeta(\phi^{n+1}) c_{\Gamma}^{n+1} - \zeta(\phi^{n}) c_{\Gamma}^{n} }{ \Delta t } +  \boldsymbol{\nabla} \cdot ( \zeta(\phi^{n}) c_{\Gamma}^{n} \boldsymbol{u}^{n} )  = \frac{1}{Pe_{\Gamma}}  \boldsymbol{\nabla} \cdot \left( \zeta(\phi^{n+1}) \boldsymbol{q_{c}}^{n+1} \right) \\ 
&\boldsymbol{q_{c}}^{n+1} = m(c_{\Gamma}^{n}) \boldsymbol{\nabla} \omega_{c}^{n+1}  \\
\label{omegac-method}
&\zeta(\phi^{n+1})\omega_c^{n+1}=\frac{1}{Ca}\frac{{\varepsilon_\phi}^{-1}}{2}{\kappa_B}^{\prime}(c_{\Gamma}^{n}) \left({\omega_{\phi}^{n+1}}\right)^{2} + {Cs_{\Gamma}}{\sigma_S}^{\prime}(c_{\Gamma}^{n}) {\delta (\phi^{n+1})} \\ 
&\quad\quad\quad\quad\quad\quad~ +{Cn_{\Gamma}} \sigma_{L} \zeta(\phi^{n+1})\left({\eta_c}  \frac{f^{\prime}(c_{\Gamma}^{n+1})}{\varepsilon_c}-{\eta_c} {\varepsilon_c} \Delta c_{\Gamma}^{n+1}\right)- {Cn_{\Gamma}}\sigma_{L} {\eta_c} {\varepsilon_c} \boldsymbol{\nabla} \zeta(\phi^{n+1}) \cdot \boldsymbol{\nabla}{c_{\Gamma}^{n}}. \nonumber
\end{align} 
Here, the nonlinear term $f^{\prime}(c_{\Gamma}^{n+1})$ is likewise linearized using a first-order Taylor expansion.

3: Finally, we solve the NS equations \eqref{ns_1}--\eqref{ns_2} for the background fluid together with the local inextensibility constraint \eqref{inextend} using a projection method, which includes the following sub-steps:

Step 1. (Temporary velocity prediction) We first solve for the temporary velocity $\boldsymbol{\tilde{u}}^{n+1}$:
\begin{align}
&\rho \left( \frac{ \boldsymbol{\tilde{u}}^{n+1} - \boldsymbol{u}^{n} }{ \Delta t } \right) + \rho \boldsymbol{\nabla} \cdot \left( \boldsymbol{u}^{n} \otimes \boldsymbol{u}^{n} \right)  = \frac{1}{Re}\boldsymbol{\nabla} \cdot \left( \mu(\phi^{n+1}) \boldsymbol{D}(\boldsymbol{\tilde{u}}^{n+1}) \right) -\boldsymbol{\nabla} p^{n} + \frac{1}{Re} \boldsymbol{F}^{n} + \boldsymbol{\nabla} \cdot \left( \zeta(\phi^{n+1}) \boldsymbol{P}^{n} \lambda_{local}^{n} \right), \\
&\boldsymbol{F}^{n} = g^{n+1} \boldsymbol{\nabla} \phi^{n+1} - \zeta(\phi^{n+1}) c_{\Gamma}^{n+1} \boldsymbol{\nabla} \omega_{c}^{n+1},
\end{align}
where $\mu(\phi^{n+1})=(1+\phi^{n+1})\mu_{{in}}/2+(1-\phi^{n+1})\mu_{{out}}/2$. The fields $\phi^{n+1}$, $g^{n+1}$, $c_\Gamma^{n+1}$ and $\omega_c^{n+1}$ are obtained from \eqref{phi-method}--\eqref{omegac-method}.

Step 2. (Pressure correction) We then solve for the pressure and then update the velocity field. Having in mind of the incompressibility constraint:
\begin{align}\label{imcompress-method}
\boldsymbol{\nabla} \cdot \boldsymbol{u}^{n+1} = 0,
\end{align}
the velocity correction is given by:
\begin{align}\label{u-method}
\boldsymbol{u}^{n+1} = \boldsymbol{\tilde{u}}^{n+1} - \frac{ \Delta t }{ \rho } \boldsymbol{\nabla} \tilde{p}^{n+1}.
\end{align}

By combining \eqref{imcompress-method} and \eqref{u-method}, the pressure $\tilde{p}^{n+1}$ is obtained by solving the following Poisson equation:
\begin{align}
\nabla^{2} \tilde{p}^{n+1} = \frac{ \rho }{ \Delta t } \boldsymbol{\nabla} \cdot \boldsymbol{\tilde{u}}^{n+1}.
\end{align}

Step 3. (Velocity and pressure update) The velocity $\boldsymbol{u}^{n+1}$ and pressure ${p}^{n+1}$ are then updated as:
\begin{align}
&\boldsymbol{u}^{n+1} = \boldsymbol{\tilde{u}}^{n+1} - \frac{ \Delta t }{ \rho } \boldsymbol{\nabla} \tilde{p}^{n+1}, \\
&p^{n+1} = p^{n} + \tilde{p}^{n+1}.
\end{align}

Step 4. (Local inextensibility correction) Finally, the Lagrange multiplier $\lambda_{local}^{n+1}$ is then updated by the following method:
\begin{align}
\xi \varepsilon_{\phi}^{2} \boldsymbol{\nabla} \cdot \left( (\phi^{n+1})^{2} \boldsymbol{\nabla} \lambda_{local}^{n+1} \right) + \zeta(\phi^{n+1}) \boldsymbol{P}^{n+1} : \boldsymbol{\nabla} \boldsymbol{u}^{n} = 0.
\end{align}
where $\boldsymbol{P}^{n+1}=\boldsymbol{I} - (\boldsymbol{\nabla}\phi^{n+1}/|\boldsymbol{\nabla}\phi^{n+1}|)\otimes (\boldsymbol{\nabla}\phi^{n+1}/|\boldsymbol{\nabla}\phi^{n+1}|)$.

In addition, Dirichlet boundary conditions can imposed on $\boldsymbol{\tilde{u}}$, consistent with $\boldsymbol{u}$ in \eqref{bcu}, while no-flux boundary conditions are applied to $\phi$, $\boldsymbol{q}_\phi$, $c_\Gamma$, and $\boldsymbol{q}_c$ as in \eqref{bcphi}--\eqref{bcc}.

\bibliographystyle{unsrt}  
\bibliography{main}  
%%% Remove comment to use the external .bib file (using bibtex).


\end{document}